\documentclass[aps,pra,reprint,nofootinbib,floatfix,superscriptaddress,fleqn]{revtex4-2}
\usepackage{mathtools,amssymb,amsthm,bm,bbm,xcolor,mathdots,stmaryrd}
\usepackage[colorlinks=true, linkcolor=blue, citecolor=magenta, urlcolor=blue]{hyperref}
\pdfoutput=1

\usepackage[T1]{fontenc}
\usepackage{lmodern}
\usepackage[utf8]{inputenc}
\usepackage{microtype}
\usepackage{orcidlink}
\usepackage[caption=false]{subfig}
\usepackage{graphicx}

\theoremstyle{remark}
\newtheorem{rmk}{Remark}

\newcommand{\HH}{\mathcal{H}}
\newcommand{\PP}{\mathcal{P}}
\renewcommand{\AA}{\mathcal{A}}
\newcommand{\JJ}{\mathcal{J}}
\renewcommand{\SS}{\mathcal{S}}
\newcommand{\tr}{\operatorname{tr}}
\renewcommand{\d}{\mathrm{d}}
\newcommand{\bra}[1]{\langle #1|}
\newcommand{\ket}[1]{|#1\rangle}
\newcommand{\braket}[2]{\langle #1 | #2 \rangle}
\newcommand{\ketbra}[2]{|#1 \rangle\langle #2|}
\newcommand{\dt}{\mathrm{d}t}

\newcommand{\tauMT}{\tau_\textsc{mt}}
\newcommand{\tauML}{\tau_\textsc{ml}}
\newcommand{\dist}{\operatorname{dist}}

\newcommand{\emax}{\epsilon_{\mathrm{max}}}

\usepackage[margin=1.73cm]{geometry}

\begin{document}
\title{Margolus-Levitin quantum speed limit for an arbitrary fidelity}
\author{Niklas H{\"o}rnedal\,\orcidlink{0000-0002-2005-8694}\,}
\affiliation{Department of Physics and Materials Science, University of Luxembourg, L-1511 Luxembourg, Luxembourg}
\author{Ole S{\"o}nnerborn\,\orcidlink{0000-0002-1726-4892}\,}
\email{ole.sonnerborn@kau.se}
\affiliation{Department of Mathematics and Computer Science, Karlstad University, 651 88 Karlstad, Sweden}
\affiliation{Department of Physics, Stockholm University, 106 91 Stockholm, Sweden}

\begin{abstract}
The Mandelstam-Tamm and Margolus-Levitin quantum speed limits are two well-known evolution time estimates for isolated quantum systems. These bounds are usually formulated for fully distinguishable initial and final states, but both have tight extensions to systems that evolve between states with an arbitrary fidelity. However, the foundations of these extensions differ in some essential respects. The extended Mandelstam-Tamm quantum speed limit has been proven analytically and has a clear geometric interpretation. Furthermore, which systems saturate the limit is known. The derivation of the extended Margolus-Levitin quantum speed limit, on the other hand, is based on numerical estimates. Moreover, the limit lacks a geometric interpretation, and no complete characterization of the systems reaching it exists. In this paper, we derive the extended Margolus-Levitin quantum speed limit analytically and describe the systems that saturate the limit in detail. We also provide the limit with a symplectic-geometric interpretation, which indicates that it is of a different character than most existing quantum speed limits. At the end of the paper, we analyze the maximum of the extended Mandelstam-Tamm and Margolus-Levitin quantum speed limits and derive a dual version of the extended Margolus-Levitin quantum speed limit. The maximum limit is tight regardless of the fidelity of the initial and final states. However, the conditions under which the maximum limit is saturated differ depending on whether or not the initial state and the final state are fully distinguishable. The dual limit is also tight and follows from a time reversal argument. We describe the systems that saturate the dual quantum speed limit. 
\end{abstract}
\date{\today}
\maketitle

\section{Introduction}
A quantum speed limit (QSL) is a lower bound on the time it takes to transform a quantum system in some predetermined way. QSLs exist for all sorts of transformations of both open and closed systems. Many involve statistical quantities such as energy, entropy, fidelity, and purity \cite{Fr2016, DeCa2017, De2020, PiCiCeAdS-P2016, dCEgPlHu2013}.

A prominent QSL is attributed to Mandelstam and Tamm \cite{MaTa1945}. The Mandelstam-Tamm QSL states that the time it takes for an isolated quantum system to evolve between two fully distinguishable states is at least $\pi/2$ divided by the energy uncertainty,\footnote{In this paper, ``state'' refers to a pure quantum state unless otherwise is explicitly stated.}\textsuperscript{,}\footnote{Two states are fully distinguishable if their fidelity is zero.}\textsuperscript{,}\footnote{All quantities are expressed in units such that $\hbar=1$.}
\begin{equation}\label{eq: MT-ortho-quantum speed limit}
	\tau \geq \tauMT(0),\quad \tauMT(0)=\frac{ \pi }{ 2 \Delta H }.
\end{equation}
Mandelstam and Tamm actually showed a more general result: If the fidelity between the initial state and the final state is $\delta$, the evolution time is bounded according to 
\begin{equation}\label{eq: the Mandelstam-Tamm quantum speed limit}
	\tau \geq \tauMT(\delta),\quad \tauMT(\delta)=\frac{\arccos\sqrt{\delta} }{ \Delta H }.
\end{equation}
For $\delta=0$, estimate \eqref{eq: the Mandelstam-Tamm quantum speed limit} agrees with estimate \eqref{eq: MT-ortho-quantum speed limit}.

Anandan and Aharonov \cite{AnAh1990} provided the Mandelstam-Tamm QSL with an elegant geometric interpretation when they showed that the Fubini-Study distance between two states with fidelity $\delta$ is $\arccos\sqrt{\delta}$ and that the Fubini-Study evolution speed equals the energy uncertainty. Inequality \eqref{eq: the Mandelstam-Tamm quantum speed limit} is thus saturated for isolated systems where the state evolves along a shortest Fubini-Study geodesic. Anandan and Aharonov's work inspired the writing of Ref.\ \cite{HoAlSo2022}, which extends the Mandelstam-Tamm QSL to systems in mixed states in several different ways.

Margolus and Levitin \cite{MaLe1998} derived another QSL that is often mentioned together with Mandelstam and Tamm's. The Margolus-Levitin QSL states that the time it takes for an isolated quantum system to evolve between two fully distinguishable states is at least $\pi/2$ divided by the expected energy shifted by the smallest energy,
\begin{equation}\label{eq: ML-ortho-quantum speed limit}
	\tau \geq \tauML(0),\quad \tauML(0)=\frac{ \pi }{ 2 \langle H - \epsilon_0 \rangle }.
\end{equation}

Giovannetti \emph{et al.}\ \cite{GiLlMa2003} extended Margolus and Levitin's QSL to an arbitrary fidelity. More precisely, they showed that the evolution time of an isolated system is bounded from below according to
\begin{equation}\label{eq: the Margolus-Levitin quantum speed limit}
	\tau \geq \tauML(\delta),\quad \tauML(\delta)=\frac{ \alpha(\delta) }{ \langle H - \epsilon_0 \rangle },
\end{equation}
where $\alpha$ is a function that depends only on the fidelity $\delta$ between the initial and final states; see Sec.\ \ref{sec: The generalized Margolus-Levitin quantum speed limit}. Giovannetti \emph{et al.}\ also showed that \eqref{eq: the Margolus-Levitin quantum speed limit}, like \eqref{eq: the Mandelstam-Tamm quantum speed limit}, is tight, which means that for each $\delta$ there is a system for which the inequality in \eqref{eq: the Margolus-Levitin quantum speed limit} is an equality. However, apart from that, the situation is different from the case of the Mandelstam-Tamm QSL: A closed formula for $\alpha$ does not exist, the derivation of \eqref{eq: the Margolus-Levitin quantum speed limit} rests on numerical estimates, there is no classification of the systems saturating \eqref{eq: the Margolus-Levitin quantum speed limit}, and a geometric interpretation like that of \eqref{eq: the Mandelstam-Tamm quantum speed limit} is lacking \cite{Fr2016, DeCa2017, TaEsDaMa-Fi2013, Ta2014}.

In this paper, we derive the extended Margolus-Levitin QSL analytically and characterize the systems that saturate this estimate (Sec.\ \ref{sec: The generalized Margolus-Levitin quantum speed limit}). Furthermore, we show that the extended Margolus-Levitin QSL has a symplectic-geometric interpretation and is connected to the Aharonov-Anandan geometric phase \cite{AhAn1987} (Sec.\ \ref{sec: Geometry of the Margolus-Levitin quantum speed limit}). The Margolus-Levitin QSL thus differs fundamentally from most existing QSLs.  That the estimates in \eqref{eq: ML-ortho-quantum speed limit} and \eqref{eq: the Margolus-Levitin quantum speed limit} are either invalid or correct but not tight unless we shift the expected energy with the smallest energy suggests that a ground state will play a central role in the derivation of the extended Margolus-Levitin QSL and in its geometric interpretation.

The maximum of the Mandelstam-Tamm and the extended Margolus-Levitin QSL is also a QSL. The characteristics of the maximum QSL differ depending on whether or not the initial state and final state are fully distinguishable. We explain why this is so, and we provide several QSLs that are related to but less sharp than the extended Margolus-Levitin QSL. We also derive a dual version of the extended Margolus-Levitin QSL involving the largest rather than the smallest energy (Sec.\ \ref{se: related and approximated}). The paper concludes with a summary and a comment on the difficulty of extending the Margolus-Levitin quantum speed limit to driven systems (Sec.\ \ref{sec: Outlook}). For a more detailed discussion of this difficulty, see Ref.\ \cite{HoSo2023}.

\section{The extended Margolus-Levitin quantum speed limit}
\label{sec: The generalized Margolus-Levitin quantum speed limit}
\begin{figure}[t]
	\centering
	\includegraphics[width=0.97\linewidth]{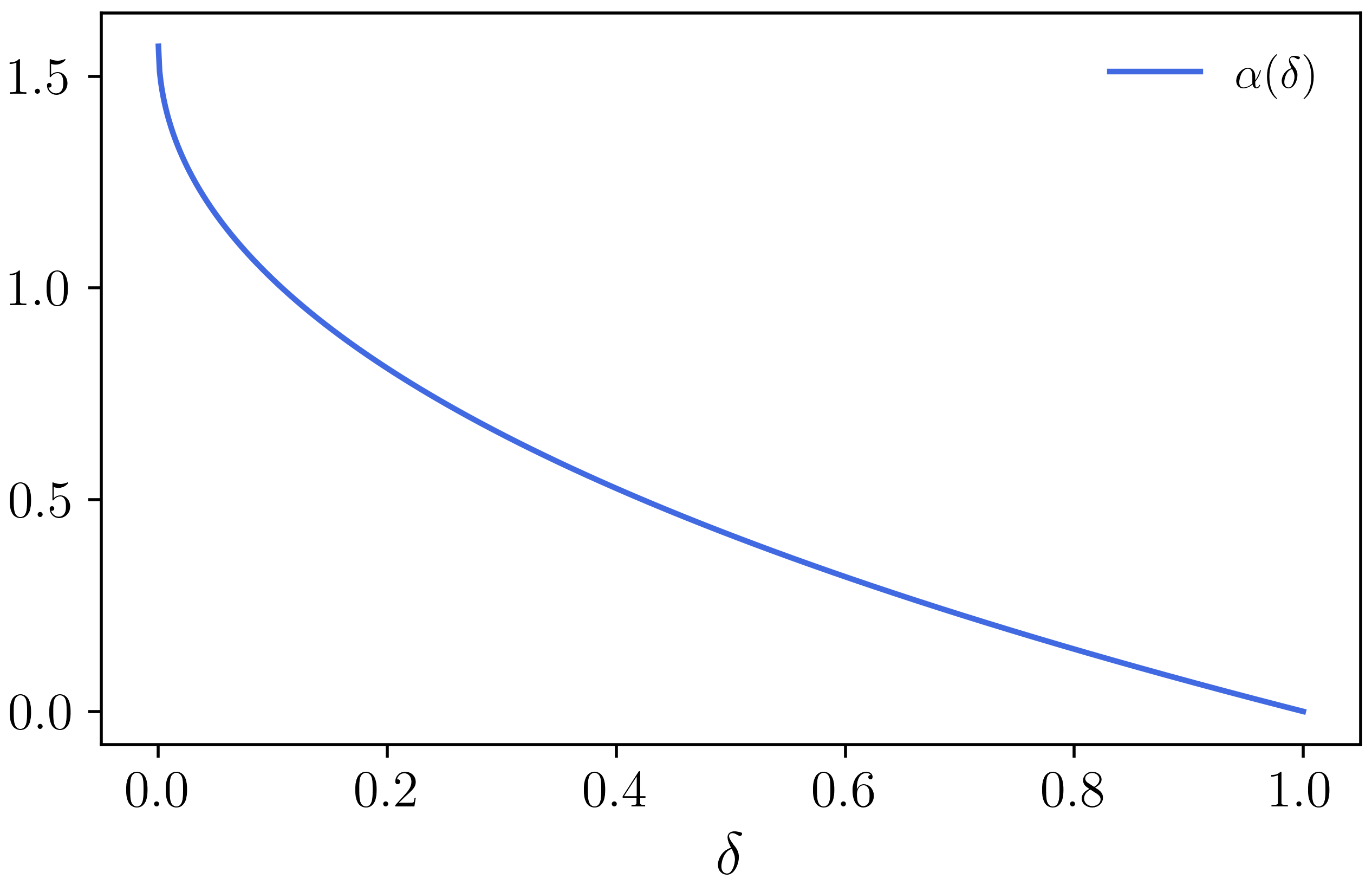}
	\caption{The graph of $\alpha$ as a function of the fidelity $\delta$ between the initial state and the final state. If $\delta=0$, the initial and final states are fully distinguishable, and $\alpha(\delta)=\pi/2$. In this case we recover the Margolus-Levitin QSL \eqref{eq: ML-ortho-quantum speed limit}. If $\delta=1$, the initial and final states are the same, and $\alpha(\delta)=0$. This is reasonable since it takes no time to remain in the initial state.}
	\label{fig: graph of alpha}
\end{figure}

The function $\alpha$ in Giovannetti \emph{et al.}'s estimate \eqref{eq: the Margolus-Levitin quantum speed limit} is\footnote{The $\alpha$ in Ref.\ \cite{GiLlMa2003} equals the $\alpha$ in \eqref{alpha} multiplied by $2/\pi$.}
\begin{equation}\label{alpha}
    \alpha(\delta)
    = \min_{z^2\leq \delta} \bigg\{\frac{1+z}{2}\arccos\bigg(\frac{2\delta-1-z^2}{1-z^2}\bigg) \bigg\}.
\end{equation}
For each $\delta$, the minimum on the right-hand side is assumed for a unique $z$. Figure \ref{fig: graph of alpha} shows the graph of $\alpha$. 
Note that $\alpha$ depends only on the fidelity $\delta$ between the initial and final states. The fidelity, or overlap, between two pure states $\rho_a$ and $\rho_b$ is $\tr(\rho_a\rho_b)$.

Although the Margolus-Levitin QSL \eqref{eq: ML-ortho-quantum speed limit} is quite surprising, its proof is relatively simple \cite{MaLe1998}. Giovannetti \emph{et al.}'s estimate \eqref{eq: the Margolus-Levitin quantum speed limit} reduces to the Margolus-Levitin QSL for $\delta=0$, but the derivation in Ref.\ \cite{GiLlMa2003} for a general $\delta$ is rather complicated. Moreover, it is partly based on numerical calculations. In this section, we derive \eqref{eq: the Margolus-Levitin quantum speed limit} analytically. We also characterize the systems, that is, the states and Hamiltonians that saturate \eqref{eq: the Margolus-Levitin quantum speed limit}.

For simplicity, we write $\langle H \rangle$ for the expected energy of a system without reference to its state. Furthermore, we write $\langle H - \epsilon_0\rangle$ for the difference between $\langle H \rangle$ and the smallest eigenvalue $\epsilon_0$ of $H$. We call this quantity the normalized expected energy. In this paper we only consider isolated systems, that is, systems where a time-independent Hamiltonian governs the dynamics. For such systems, the expected energy and normalized expected energy are conserved quantities. The expected energy and normalized expected energy thus depend on the initial state but do not change as the state evolves.

\subsection{The extended Margolus-Levitin quantum speed limit for a two-dimensional system}\label{sec: 2D}
In this section we show that Giovannetti \emph{et al.}'s estimate \eqref{eq: the Margolus-Levitin quantum speed limit} is valid and tight for qubit systems. Derivations of the statements in this section can be found in Appendix \ref{appendix 0}.

Consider a qubit system with Hamiltonian
\begin{equation}\label{eq: qubit Hamiltonian}
    H=\epsilon_0\ketbra{0}{0} + \epsilon_1\ketbra{1}{1},\quad \epsilon_0<\epsilon_1,
\end{equation}
the vectors $\ket{0}$ and $\ket{1}$ being orthonormal. We identify each qubit state $\rho$ with a unit length vector $\mathbf{r}=(x,y,z)$ called the Bloch vector of $\rho$ by defining
\begin{align}
    x &= \bra{1}\rho\ket{0} + \bra{0}\rho\ket{1}, \label{x}\\
    y &= i(\bra{1}\rho\ket{0} - \bra{0}\rho\ket{1}), \label{y}\\
    z &= 1- 2  \bra{0}\rho\ket{0}. \label{z}
\end{align}
The dynamics induced by $H$ causes the Bloch vectors to rotate about the $z$ axis with a constant inclination and an azimuthal angular speed $\epsilon_1-\epsilon_0$. Furthermore, a state's expected energy is determined by, and determines, the $z$ coordinate of its Bloch vector:
\begin{equation}\label{eq: the z-coordinate}
    z=\frac{2\langle H\rangle-\epsilon_1-\epsilon_0}{\epsilon_1-\epsilon_0}=2\frac{\langle H-\epsilon_0\rangle}{\epsilon_1-\epsilon_0}-1.
\end{equation}
Two states thus have the same expected energy if and only if their Bloch vectors have the same $z$ coordinate. 

Let $\mathbf{r}_a$ and $\mathbf{r}_b$ be the Bloch vectors of two states with a common $z$ coordinate $z$ and fidelity $\delta$, and suppose that $H$ causes $\mathbf{r}_a$ to rotate to $\mathbf{r}_b$ in time $\tau$. Using \eqref{x}--\eqref{z} one can show that the inner product between $\mathbf{r}_a$ and $\mathbf{r}_b$ relates to the fidelity $\delta$ as
\begin{equation}\label{eq: inner product and fidelity}
    \mathbf{r}_a \cdot \mathbf{r}_b = 2\delta - 1.
\end{equation}
To determine the distance traveled by the rotating Bloch vector consider the orthogonal projections $\bar{\mathbf{r}}_a$ and $\bar{\mathbf{r}}_b$ of $\mathbf{r}_a$ and $\mathbf{r}_b$ on the $xy$ plane. During the evolution, $\bar{\mathbf{r}}_a$ rotates to $\bar{\mathbf{r}}_b$ along the peripheral arc of a circular sector in the $xy$ plane of radius $|\bar{\mathbf{r}}_a|=(1-z^2)^{1/2}$ and apex angle $\arccos(\bar{\mathbf{r}}_a\cdot \bar{\mathbf{r}}_b/ |\bar{\mathbf{r}}_a|^2)=\arccos[(2\delta-1-z^2)/(1-z^2)]$. The 
distance traveled by the rotating Bloch vector equals the length of the peripheral arc and is, thus,
\begin{equation}\label{tolv}
    \sqrt{1-z^2}\arccos\left(\frac{2\delta-1-z^2}{1-z^2}\right).
\end{equation}
The speed of the rotating Bloch vector is $(\epsilon_1-\epsilon_0)(1-z^2)^{1/2}$, and hence, by \eqref{tolv}, the evolution time is
\begin{equation}\label{tretton}
	\tau=\frac{1}{\epsilon_1-\epsilon_0}\arccos\left(\frac{2\delta-1-z^2}{1-z^2}\right).
\end{equation}
Combined with \eqref{eq: the z-coordinate}, this gives that
\begin{equation}\label{eq: upper bound}
    \tau \langle H - \epsilon_0\rangle = \frac{1+z}{2}\arccos\bigg(\frac{2\delta-1-z^2}{1-z^2}\bigg).
\end{equation}

The relation \eqref{eq: inner product and fidelity} implies that the $z$ coordinate squared of the Bloch vectors of two states with the same expected energy is less than the fidelity between the states:
\begin{equation}
	2\delta - 1
	= \mathbf{r}_a \cdot \mathbf{r}_b
	\geq 
	2z^2 - \mathbf{r}_a \cdot \mathbf{r}_a
	= 2z^2-1.
\end{equation}
Conversely, each expected energy level corresponding to a $z$ such that $z^2\leq\delta$ contains Bloch vectors of states with fidelity $\delta$. We conclude that $\tau\geq\tauML(\delta)$ for a qubit. Equality holds if and only if the $z$ coordinate of the Bloch vector of the initial state minimizes the right side of \eqref{eq: upper bound} over the interval $-\sqrt{\delta}\leq z\leq\sqrt{\delta}$ and thus is such that
\begin{equation}\label{zeta}
	\frac{1+z}{2}\arccos\bigg(\frac{2\delta-1-z^2}{1-z^2}\bigg)=\alpha(\delta).
\end{equation}

\subsection{The extended Margolus-Levitin quantum speed limit for systems of arbitrary dimension}
Section \ref{sec: 2D} shows that Giovannetti \emph{et al.}'s estimate \eqref{eq: the Margolus-Levitin quantum speed limit} is valid and tight for qubit systems. Section \ref{sec: 2D} also shows that for an arbitrary isolated system with Hamiltonian $H$ there is a state that evolves into one with fidelity $\delta$ in such a way that \eqref{eq: the Margolus-Levitin quantum speed limit} is an identity: Let $\ket{0}$ and $\ket{1}$ be eigenvectors of $H$ with eigenvalues $\epsilon_0$ and $\epsilon_1$, with $\epsilon_1 > \epsilon_0$. Choose a $\rho$ with support in the span of $\ket{0}$ and $\ket{1}$ and such that $z$ defined by \eqref{z} satisfies \eqref{zeta}. Then $\rho$ will evolve to a state with fidelity $\delta$ in time $\tau=\tauML(\delta)$. Conversely, for any system in a state $\rho$, a Hamiltonian exists that transforms $\rho$ into a state with fidelity $\delta$ in such a way that \eqref{eq: the Margolus-Levitin quantum speed limit} is saturated: Take an $H$ whose sum of two eigenspaces, one of which corresponds to its smallest eigenvalue $\epsilon_0$, contains the support of $\rho$. Adjust $H$'s spectrum so that $z$ defined by \eqref{z} satisfies \eqref{zeta}. Then $\rho$ will evolve to a state with fidelity $\delta$ in time $\tau=\tauML(\delta)$. 

An \emph{effective qubit} for $H$ is a state with support in the sum of two eigenspaces of $H$. It behaves like a genuine qubit in that its support evolves in the linear span of two energy eigenvectors, one from each eigenspace covering the initial support. The above discussion shows that \eqref{eq: the Margolus-Levitin quantum speed limit} holds for and can always be saturated by an effective qubit. We say that a state is \emph{partly grounded} if the eigenspace corresponding to $\epsilon_0$ is not contained in the kernel of the state or, equivalently, if the eigenspace of $\epsilon_0$ is not orthogonal to the support of the state. In Appendix \ref{appendix A} we show the first main result of this paper:
%
If $\tau \langle H - \epsilon_0\rangle$ assumes its smallest possible value when evaluated for all Hamiltonians $H$, states $\rho$, and $\tau\geq 0$ such that $H$ transforms $\rho$ into a state with fidelity $\delta$ in time $\tau$, then $\rho$ is a partly grounded effective qubit for $H$ and $\tau \langle H - \epsilon_0\rangle = \alpha(\delta)$.
%
The extended Margolus-Levitin QSL \eqref{eq: the Margolus-Levitin quantum speed limit} thus holds generally and is a tight estimate saturable in all dimensions. We interpret \eqref{eq: the Margolus-Levitin quantum speed limit} geometrically in Sec.\ \ref{sec: Geometry of the Margolus-Levitin quantum speed limit}. There we see, among other things, that one can interpret $\alpha(\delta)$ as an extremal dynamical phase. Notice the contrast with the Mandelstam-Tamm QSL, where the numerator is a geodesic distance.

\begin{rmk}\label{rmk: ett}
If we select a subset of the spectrum of $H$ and consider only initial states with support in the sum of the eigenspaces of the eigenvalues in the subset, then the support of the evolved state will remain in that sum. The proof in Appendix \ref{appendix A} shows that the evolution time of each such state satisfies the inequality
\begin{equation}\label{eq: generalized extended ML-quantum speed limit}
	\tau \geq \frac{\alpha(\delta)}{\langle H - \epsilon_0' \rangle},
\end{equation}
with $\epsilon_0'$ being the smallest eigenvalue in the subset.\footnote{To avoid having to treat trivial cases separately, we assume that the subset contains at least two different eigenvalues.} Furthermore, by Sec.\ \ref{sec: 2D}, inequality  \eqref{eq: generalized extended ML-quantum speed limit} can be saturated with an effective qubit that evolves in the sum of the eigenspaces corresponding to two eigenvalues in the subset, one of which is $\epsilon_0'$. As a special case, we have that the evolution time is bounded according to 
\begin{equation}
	\tau \geq \frac{\alpha(\delta)}{\langle H - \epsilon_0'' \,\rangle },
\end{equation}
where $\epsilon_0''$ is the smallest occupied energy, that is, the smallest eigenvalue of $H$ whose corresponding eigenspace is not annihilated by $\rho$. Often, the Margolus-Levitin QSL is formulated with the expected energy shifted by the smallest occupied energy rather than the smallest energy. Mathematically, however, there is no difference because we can always reduce the Hilbert space to an effective Hilbert space and consider the smallest occupied energy as the smallest energy. (However, see Remark \ref{rmk: tva}.)
\end{rmk}

\begin{rmk}\label{rmk: tva}
The state must evolve in the span of two eigenvectors of $H$ to saturate \eqref{eq: the Margolus-Levitin quantum speed limit}, one of which has eigenvalue $\epsilon_0$. No requirements are placed on the eigenvalue $\epsilon_1$ of the second eigenvector except that it must differ from $\epsilon_0$. However, if we want the evolution time to be as short as possible, $\epsilon_1$ must be the largest eigenvalue of $H$. This follows from \eqref{eq: the z-coordinate} since saturation of \eqref{eq: the Margolus-Levitin quantum speed limit} implies that the quotient $\langle H - \epsilon_0\rangle / (\epsilon_1-\epsilon_0)$ is independent of $\epsilon_1$. Thus, the maximum value of $\langle H - \epsilon_0\rangle$, and consequently the minimum value of $\tau$, is obtained for the $\epsilon_1$ maximizing the difference $\epsilon_1-\epsilon_0$. The observation that the state that saturates \eqref{eq: the Margolus-Levitin quantum speed limit} with the shortest possible evolution time is an effective qubit with support in the sum of the eigenspaces belonging to the largest and the smallest eigenvalue generalizes the main result in Ref.\ \cite{SoBjTsTr1999} to arbitrary fidelity; see also Ref.\ \cite{LeTo2009}. A corresponding statement holds if we restrict the Hilbert space as in Remark \ref{rmk: ett}.
\end{rmk}

In contrast to the energy uncertainty, we cannot consider the normalized expected energy $\langle H - \epsilon_{0} \rangle$ as a measure of a state's rate of change. Since for each state there exist Hamiltonians $H_1$ and $H_2$, with smallest eigenvalues $\epsilon_0^1$ and $\epsilon_0^2$, that identically evolve the state but for which $\langle H_1 - \epsilon_0^1 \rangle$ and $\langle H_2 - \epsilon_0^2 \rangle$ are different. Thus, unlike most QSLs, the extended Margolus-Levitin QSL is not a quotient of a distance and a speed \cite{Fr2016, DeCa2017, TaEsDaMa-Fi2013, Ta2014}.

\section{Geometry of the extended Margolus-Levitin quantum speed limit}
\label{sec: Geometry of the Margolus-Levitin quantum speed limit}

\begin{figure}[t]
	\centering
	\includegraphics[width=0.7\linewidth]{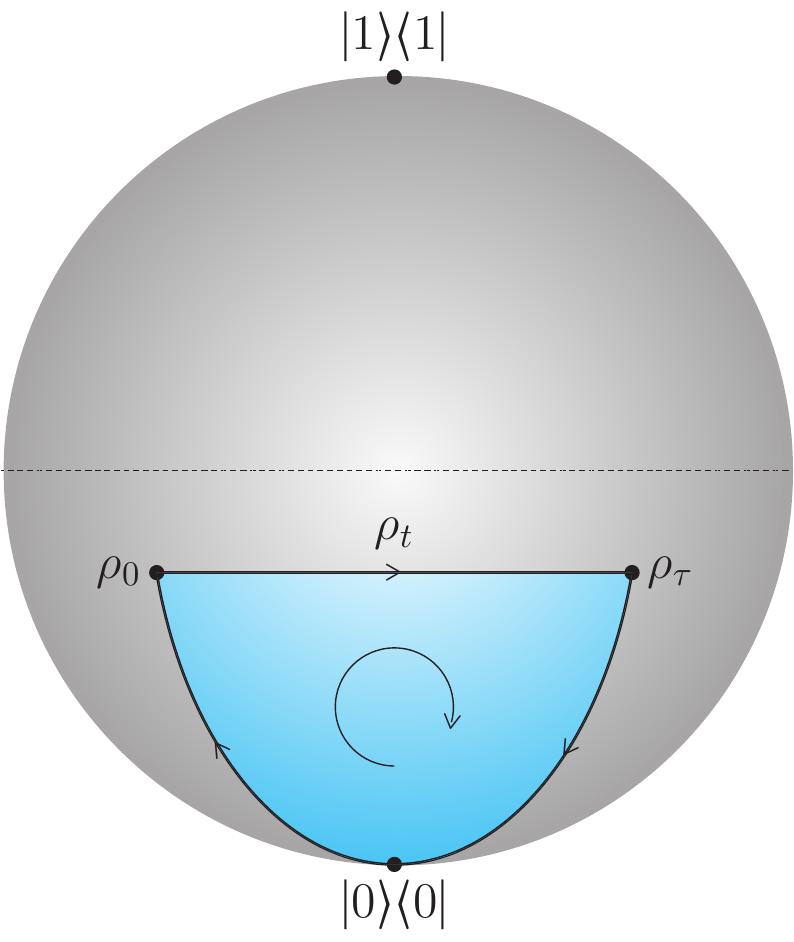}
	\caption{An oriented surface in the Bloch sphere bounded by the evolution curve $\rho_t$ and the shortest geodesics connecting the initial and final states $\rho_0$ and $\rho_\tau$ to the ground state $\ketbra{0}{0}$. The orientation of the surface is such that its symplectic area is negative. The negative of the symplectic area is minimal if and only if the extended Margolus-Levitin QSL is saturated.}
	\label{fig: Symplectic area in the Bloch sphere}
\end{figure}

Equations \eqref{x}--\eqref{z} describe a diffeomorphism between the projective Hilbert space of qubit states and the Bloch sphere. (Thus, we can identify qubit states with their corresponding Bloch vectors.) We push forward the Fubini-Study Riemannian metric and symplectic form using this diffeomorphism.\footnote{The Fubini-Study distance is equal to half of the standard distance function on the unit sphere, and the Fubini-Study symplectic form is equal to half the standard area form on the unit sphere.} The expression in \eqref{eq: upper bound} is then the negative of the symplectic area of a surface with a triangular boundary in the Bloch sphere. The path traced out by the evolving state and the shortest geodesics connecting the initial and final state to the lowest energy state $\ketbra{0}{0}$ form the boundary of the surface; see Fig.\ \ref{fig: Symplectic area in the Bloch sphere}.
Notice that we have oriented the boundary so that the surface has the reverse orientation compared with the standard orientation of the Bloch sphere. In Sec.\ \ref{del A}, we show that $\tau\langle H - \epsilon_{0} \rangle$ is equal to the negative of the symplectic area of such a triangular surface also in the general case.

The expected energy level to which the initial qubit state $\rho$ belongs is a geodesic sphere centered at $\ketbra{0}{0}$, that is, a sphere made up of all states at a fixed distance from $\ketbra{0}{0}$. The radius of the geodesic sphere is
\begin{equation}
	r=\arccos\sqrt{\bra{0}\rho\ket{0}}=\arccos\sqrt{(1-z)/2}.
\end{equation}
Therefore, $z=-\cos(2r)$. If we substitute $z$ for $-\cos(2r)$ on the right-hand side of \eqref{alpha} we get
\begin{equation}\label{eq: alpha in terms of r}
	\alpha(\delta)
	= \min_{r}\bigg\{\sin^2r \arccos\bigg(1-\frac{2(1-\delta)}{\sin^2(2r)}\bigg)\bigg\},
\end{equation}
where $r$ ranges from $\frac{1}{2} \arccos \sqrt{\delta}$ to $\frac{1}{2} \arccos(-\sqrt{\delta})$. Section \ref{del B} shows that the expression minimized on the right-hand side is an extremal dynamical phase in a gauge specified by a stationary state with eigenvalue $\epsilon_0$. First, in Sec.\ \ref{sec: null}, we interpret $\tau\langle H-\epsilon_0\rangle$ as a dynamical phase.

\subsection{Evolution time times normalized expected energy as a dynamical phase}\label{sec: null}

Consider a quantum system modeled on a finite-dimensional Hilbert space $\HH$. Let $\SS$ be the unit sphere in $\HH$ and $\PP$ be the projective Hilbert space of orthogonal projection operators of rank $1$ on $\HH$.\footnote{Such operators represent pure states.} The Hopf bundle is the $U(1)$-principal bundle $\eta$ that sends each $\ket{\psi}$ in $\SS$ to the corresponding state $\ketbra{\psi}{\psi}$ in $\PP$. The Berry connection on the Hopf bundle is defined as 
\begin{equation}
    \AA\ket{\dot\psi}=i\braket{\psi}{\dot\psi}
\end{equation}
on tangent vectors $\ket{\dot\psi}$ at $\ket{\psi}$.

\begin{figure}[t]
	\centering
	\includegraphics[width=0.90\linewidth]{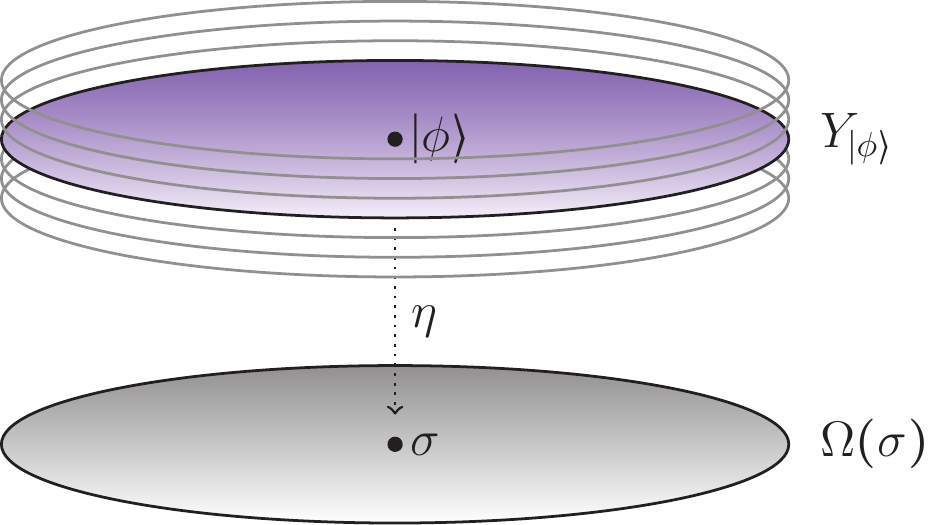}
	\caption{The preimage under the Hopf projection $\eta$ of the set $\Omega(\sigma)$ of states having nonzero fidelity with $\sigma$ is foliated by hypersurfaces $Y_{\ket{\phi}}$; one for each $\ket{\phi}$ in $\SS$ being projected to $\sigma$. The hypersurface $Y_{\ket{\phi}}$ consists of all the $\ket{\psi}$ in $\SS$ that are in phase with $\ket{\phi}$. The gauge group permutes the hypersurfaces, and $\eta$ maps each hypersurface diffeomorphically onto $\Omega(\sigma)$. We define a potential $A^\sigma$ on $\Omega(\sigma)$ by pushing down the restriction of $\AA$ to one of the hypersurfaces. The potential depends on $\sigma$ but not on the choice of hypersurface.}
	\label{fig: foliate}
\end{figure}

Assume the system has a Hamiltonian $H$. Choose an eigenstate $\sigma$ of $H$ with eigenvalue $\epsilon$. Let $\Omega(\sigma)$ be the open neighborhood of $\sigma$ consisting of all states that are not fully distinguishable from $\sigma$. Hypersurfaces $Y_{\ket{\phi}}$ in $\SS$, one for each vector $\ket{\phi}$ in the fiber over $\sigma$, foliate the preimage of $\Omega(\sigma)$ under $\eta$; see Fig.\ \ref{fig: foliate} and Ref.\ \cite{MuSi1993}. 
The hypersurface $Y_{\ket{\phi}}$ consists of all vectors $\ket{\psi}$ in $\SS$ that are in phase with $\ket{\phi}$, that is, are such that $\braket{\phi}{\psi}>0$. The gauge group permutes the hypersurfaces, and $\eta$ maps each hypersurface diffeomorphically onto $\Omega(\sigma)$. We can thus define a gauge potential $A^\sigma$ on $\Omega(\sigma)$ by pushing down the restriction of $\AA$ to an arbitrary hypersurface $Y_{\ket{\phi}}$. The potential depends on $\sigma$ but not on the choice of $\ket{\phi}$ in the fiber over $\sigma$.

The second main result in the paper reads: If $\rho$ is a state in $\Omega(\sigma)$, and $\rho_t$, where $0\leq t\leq\tau$, is the evolution curve starting from $\rho$, then $\rho_t$ is contained in $\Omega(\sigma)$ and
\begin{equation}\label{eq: dynamical phase}
    \tau\langle H-\epsilon\rangle = \int_{\rho_t} A^\sigma.
\end{equation}
In other words, $\tau\langle H-\epsilon\rangle$ is the dynamical phase of $\rho_t$ in a gauge associated with $\sigma$.\footnote{Some authors call the negative of the right-hand side of \eqref{eq: dynamical phase} the dynamical phase of $\rho_t$.} To prove \eqref{eq: dynamical phase} select a $\ket{\phi}$ in the fiber over $\sigma$, let $\ket{\psi}$ be the vector over $\rho$ in phase with $\ket{\phi}$, and let $\ket{\psi_t}$ be the curve that extends from $\ket{\psi}$ and has the velocity field $\ket{\dot\psi_t}=-i(H-\epsilon)\ket{\psi_t}$. The curve $\ket{\psi_t}$ is in phase with $\ket{\phi}$ and projects to $\rho_t$:
\begin{align}
    & \braket{\phi}{\psi_t} 
    = \bra{\phi} e^{-it(H-\epsilon)}\ket{\psi}
    = \braket{\phi}{\psi} 
    > 0, \\
    & \ketbra{\psi_t}{\psi_t}
    = e^{-it(H-\epsilon)} \ketbra{\psi}{\psi}e^{it(H-\epsilon)}
    = \rho_t.
\end{align}
Hence, 
\begin{equation}\label{tjutre}
    \int_{\rho_t} A^\sigma
    =\int_{\ket{\psi_t}}\AA
    =\tau\langle H-\epsilon\rangle.
\end{equation}
Notice that $\epsilon = \epsilon_0$ if $\sigma$ is a ground state. 

\subsection{Evolution time times normalized expected energy as the negative of a symplectic area}\label{del A}
We equip the projective Hilbert space $\PP$ with the Fubini-Study Riemannian metric $g$ and symplectic form $\omega$. For tangent vectors $\dot\rho_a$ and $\dot\rho_b$ at $\rho$,\footnote{We normalize $g$ and $\omega$ asymmetrically as this gives rise to cleaner formulas.}
\begin{align}
    &g(\dot\rho_a,\dot\rho_b)=\frac{1}{2}\tr(\dot\rho_a\dot\rho_b),\label{metric} \\
    &\omega(\dot\rho_a,\dot\rho_b)= -i\tr([\dot\rho_a,\dot\rho_b]\rho)\label{symplectic form}.
\end{align}
The geodesic distance function associated with the Fubini-Study metric is
\begin{equation}\label{distance}
    \dist(\rho_a,\rho_b) = \arccos\sqrt{\tr(\rho_a\rho_b)}.
\end{equation}

\begin{figure}[t]
    \centering
    \includegraphics[width=\linewidth]{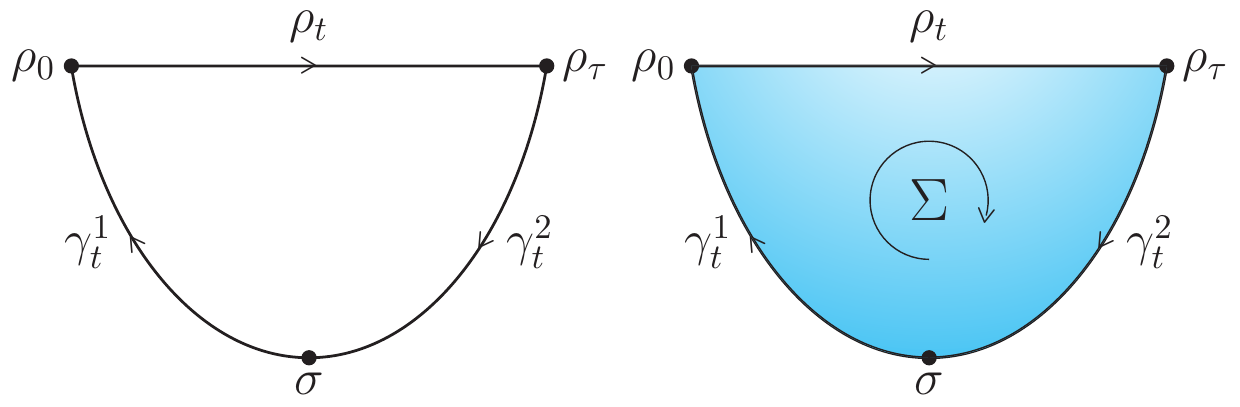} 
    \caption{On the left, the $\sigma$ closure $\rho_t^\sigma$ made up of the shortest geodesic $\gamma^1_t$ from $\sigma$ to the initial state $\rho_0$, the evolution curve $\rho_t$, and the shortest geodesic $\gamma^2_t$ from the final state $\rho_\tau$ to $\sigma$. On the right, a Seifert surface $\Sigma$ for $\rho_t^\sigma$. The circular arrow indicates the orientation of $\Sigma$ compatible with that of $\rho_t^\sigma$.}
    \label{fig: Completion}
\end{figure}

Suppose the system is in a state $\rho$ and a Hamiltonian $H$ governs its dynamics. Let $\rho_t$, $0\leq t\leq \tau$, represent the evolving state. Since $H$ is time-independent, the distances between $\rho_t$ and the eigenstates of $H$ are preserved. Let $\sigma$ be an eigenstate with eigenvalue $\epsilon$ located at a distance $r < \pi/2$ from $\rho$. Furthermore, let $\gamma_t^1$ and $\gamma_t^2$ be the shortest unit speed geodesics from $\sigma$ to $\rho_0$ and from $\rho_\tau$ to $\sigma$, respectively. The $\sigma$ closure of $\rho_t$ is the concatenation $\rho^\sigma_t=\gamma^1_t\ast\rho_t\ast\gamma^2_t$ defined as 
\begin{equation}\label{eq: the concatenation}
    \rho^\sigma_t
    =\begin{cases}
    \gamma^1_{t} & \text{for}\quad 0\leq t\leq r, \\
    \rho_{t-r} & \text{for}\quad r\leq t\leq \tau + r, \\
    \gamma^2_{t-\tau-r}& \text{for}\quad \tau+r\leq t\leq \tau+2r.
    \end{cases}
\end{equation}
The left part of Fig.\ \ref{fig: Completion} illustrates the $\sigma$ closure.
Below we show the third main result of the paper:
\begin{equation}\label{eq: result 1}
    \tau\langle H-\epsilon\rangle
 		= -\iint_\Sigma\omega,
\end{equation} 
where $\Sigma$ is any Seifert surface for $\rho^\sigma_t$ that is homologous to a Seifert surface for $\rho^\sigma_t$ in $\Omega(\sigma)$. A Seifert surface for $\rho^\sigma_t$ is an oriented surface $\Sigma$ in $\PP$ whose boundary is parametrized by $\rho^\sigma_t$ as illustrated in the right part of Fig.\ \ref{fig: Completion}. For example, the ruled surface obtained by connecting $\sigma$ and each $\rho_t$ with the shortest arclength-parametrized geodesic is such a Seifert surface. In Appendix \ref{appendix B} we provide formulas for the geodesics $\gamma_t^1$ and $\gamma_t^2$ and explicitly construct a ruled Seifert surface for $\rho^\sigma_t$.

The pull-back of $\omega$ to $\SS$ by the Hopf projection is exact and equals the negative of the Berry curvature, $\eta^*\omega=-\d\AA$ \cite{MuSi1993}. We construct a lift $\ket{\psi_t^\sigma}$ of $\rho_t^\sigma$ to $\SS$ as follows. Let $\ket{\phi}$ be any vector in the fiber over $\sigma$, let $\ket{\psi_t}$ be the lift of $\rho_t$ which is in phase with $\ket{\phi}$, and for $0\leq t\leq r$ define $\ket{\phi^1_t}$ and $\ket{\phi^2_t}$ as
\begin{align}
	&\hspace{-17pt}\ket{\phi^1_t}=\cos t\ket{\phi}+\frac{\sin t}{\sin r}\big(\ket{\psi_0} - \cos r\ket{\phi}\big),\label{eq: lift 1} \\
	&\hspace{-17pt}\ket{\phi^2_t}=\cos (r-t)\ket{\phi}+\frac{\sin (r-t)}{\sin r}\big(\ket{\psi_\tau} - \cos r\ket{\phi}\big).\label{eq: lift 2}
\end{align}
The curves $\ket{\phi^1_t}$ and $\ket{\phi^2_t}$ are lifts of $\gamma_t^1$ and $\gamma_t^2$ connecting $\ket{\phi}$ to $\ket{\psi_0}$ and $\ket{\psi_\tau}$ to $\ket{\phi}$, respectively; see Appendix \ref{appendix B}. We define $\ket{\psi_t^\sigma}$ as the concatenation $\ket{\phi^1_t}\ast \ket{\psi_t}\ast \ket{\phi^2_t}$. The curve $\ket{\psi_t^\sigma}$ is a lift of $\rho_t^\sigma$ that is in phase with $\ket{\phi}$.

Suppose that $\Sigma$ is homologous to a Seifert surface $\Sigma'$ in $\Omega(\sigma)$. The Hopf bundle restricts to a diffeomorphism from the hypersurface $Y_{\ket{\phi}}$ of vectors in $\SS$ that are in phase with $\ket{\phi}$ onto $\Omega(\sigma)$ \cite{MuSi1993}. Lift $\Sigma'$ to a Seifert surface for $\ket{\psi_t^\sigma}$ in $Y_{\ket{\phi}}$. By Stoke's theorem,
\begin{equation}\label{eq: Stokes theorem}
    \iint_\Sigma\omega
    = \iint_{\Sigma'}\omega
    = -\iint_{\Sigma'}\d\AA
    = -\oint_{\ket{\psi_t^\sigma}}\AA.
\end{equation}
The first identity is a consequence of $\Sigma-\Sigma'$ being a homo\-logical boundary and that $\omega$ is closed; cf.\ Remark \ref{rmk: fem} below. The second identity results from $\eta$ being a diffeomorphism from $Y_{\ket{\phi}}$ onto $\Omega(\sigma)$ and $-\d\AA$ being the pull-back of $\omega$. The third identity follows from $\ket{\psi_t^\sigma}$ parametrizing the boundary of the lift of $\Sigma'$. 

A direct calculation shows that the Berry connection annihilates the velocity fields of $\ket{\phi^1_t}$ and $\ket{\phi^2_t}$:
\begin{equation}\label{eq: Berry kills}
	\AA\ket{\dot\phi^j_t} = i\braket{\phi^j_t}{\dot\phi^j_t} = 0,\quad j=1,2.
\end{equation}
Thus we have that 
\begin{equation}\label{eq: end of the line}
    \oint_{\ket{\psi_t^\sigma}}\AA
    = \int_{\ket{\psi_t}}\AA.
\end{equation}
Equations \eqref{tjutre}, \eqref{eq: Stokes theorem}, and \eqref{eq: end of the line} yield the third main result \eqref{eq: result 1}. Note that the calculations rely on $\rho$ not being fully distinguishable from $\sigma$, cf.\ Remark \ref{rmk: ett}.

\begin{rmk}\label{rmk: fem}
The homology class of the projectivization of any two-dimensional subspace of $\HH$, that is, a Bloch sphere, generates the second singular homology group of $\PP$ with integer coefficients \cite{Br1993}. We choose such a Bloch sphere that we orient so that its symplectic area is positive. The area is $2\pi$, and $\omega/2\pi$ is thus of integral class.

The difference between two Seifert surfaces for $\rho_t^\sigma$ is a $2$-cycle. The homology class of such a difference is an integer multiple of the homology class for the Bloch sphere. It follows that the difference of the symplectic areas of two Seifert surfaces for $\rho_t^\sigma$ is an integer multiple of the symplectic area of the Bloch sphere. Hence, for an arbitrary Seifert surface $\Sigma$ for $\rho_t^\sigma$,
\begin{equation}\label{trettifyra}
	\tau\langle H-\epsilon\rangle =-\iint_\Sigma\omega\mod 2\pi.
\end{equation}
We can equivalently express this as $\tau\langle H-\epsilon\rangle$ being equal to the Aharonov-Anandan geometric phase \cite{AhAn1987} of the $\sigma$ closure of $\rho_t$ modulo $2\pi$. This connection to the Aharonov-Anandan phase is utilized in Ref.\ \cite{HoSo2023c} where QSLs for cyclic systems are derived.
\end{rmk}

\subsection{$\alpha(\delta)$ as an extremal dynamical phase}\label{del B}
Let $\sigma$ be any state and $S(\sigma,r)$ be the geodesic sphere of radius $r$ centered at $\sigma$. The geodesic sphere consists of all states at distance $r$ from $\sigma$. Fix a fidelity $\delta$ and choose $r$ such that $\arccos\sqrt{\delta}  \leq 2r \leq \arccos( - \sqrt{\delta})$. The geodesic sphere $S(\sigma,r)$ is then contained in $\Omega(\sigma)$ and includes states between which the fidelity is $\delta$. 

Write $\Gamma(\sigma,r,\delta)$ for the space of smooth curves in $S(\sigma,r)$ that extend between two states with fidelity $\delta$. Define $\JJ^\sigma_{r,\delta}$ to be the functional that assigns the dynamical phase to each $\rho_t$ in $\Gamma(\sigma,r,\delta)$ in the gauge specified by $\sigma$,
\begin{equation}
    \JJ^\sigma_{r,\delta}[\rho_t]
    = \int_{\rho_t}A^\sigma.
\end{equation}
In Appendix \ref{appendix C} we show that $\rho_t$ is an extremal for $\JJ^\sigma_{r,\delta}$ if and only if for every $\ket{\phi}$ in the fiber over $\sigma$, the lift $\ket{\psi_t}$ of $\rho_t$ which is in phase with $\ket{\phi}$ splits orthogonally as
\begin{equation}
	\ket{\psi_t} 
	= \cos r\,\ket{\phi} + \sin r\, e^{i\Lambda_t}  \ket{w}
\end{equation}
for some function $\Lambda_t$ that vanishes for $t = 0$. The corresponding extreme value is
\begin{equation}
	\JJ^\sigma_{r,\delta}[\rho_t]= \sin^2r\, \Lambda_\tau.
\end{equation}
The constraint on $\Lambda_t$ arising from the assumption that $\rho_t$ extends between two states with fidelity $\delta$ reads 
\begin{equation}
	\cos\Lambda_\tau=1-\frac{2(1-\delta)}{\sin^22r}.
\end{equation} 
Thus,
\begin{equation}\label{furti}
	\JJ^\sigma_{r,\delta}[\rho_t]
	=\pm\sin^2r\,\arccos\bigg(1-\frac{2(1-\delta)}{\sin^22r}\bigg) \mod 2\pi.
\end{equation}
According to \eqref{eq: alpha in terms of r}, $\alpha(\delta)$ is equal to the smallest positive extreme value of $\JJ^\sigma_{r,\delta}$ minimized over the interval $\frac{1}{2}\arccos\sqrt{\delta}  \leq r \leq \frac{1}{2}\arccos( - \sqrt{\delta})$. This observation is the fourth main result of the paper.

\begin{rmk}\label{rmk: sex}
If $\rho_t$ is an extremal curve for $\JJ^\sigma_{r,\delta}$, then so is $\rho_{\tau-t}$, and $\JJ^\sigma_{r,\delta}[\rho_{\tau-t}]=-\JJ^\sigma_{r,\delta}[\rho_{t}]$. Hence the $\pm$ in \eqref{furti}. This observation is related to the dual QSL derived and discussed in Sec.\ \ref{sec: the dual}.
\end{rmk}

\section{Related quantum speed limits}\label{se: related and approximated}
The maximum of the Mandelstam-Tamm and the extended Margolus-Levitin QSLs is a new QSL. Interestingly, the maximum QSL behaves differently for fully and not fully distinguishable initial and final states. We describe this difference in Sec.\ \ref{sec: the max qls}. In Sec.\ \ref{sec: the dual} we derive a QSL that extends the dual Margolus-Levitin QSL studied in Ref.\ \cite{NeAlSa2022}, and in Sec.\ \ref{sec: Approximations} we provide three QSLs that are less sharp but easier to calculate than the extended Margolus-Levitin QSL. These three QSLs are not new but can be found in the cited papers.

\subsection{The maximum quantum speed limit}\label{sec: the max qls}
According to Anandan and Aharonov \cite{AnAh1990}, the QSL of Mandelstam and Tamm \eqref{eq: the Mandelstam-Tamm quantum speed limit} is saturated if and only if the evolving state follows a shortest Fubini-Study geodesic. Furthermore, by Brody \cite{Br2003}, such a state is an effective qubit that follows the equator of the Bloch sphere associated with the two eigenvectors that have nonzero fidelity with the initial state; cf.\ Sec.\ \ref{sec: 2D}.\footnote{None of the eigenvectors need to be associated with the smallest eigenvalue of the Hamiltonian.} Levitin and Toffoli \cite{LeTo2009} showed that if we require the initial and final states to be fully distinguishable, the same holds for the Margolus-Levitin QSL \eqref{eq: ML-ortho-quantum speed limit}. However, in that case, one of the eigenvectors must belong to the smallest eigenvalue (or $\epsilon_0$ be replaced by the smallest occupied energy, cf.\ Remark \ref{rmk: ett}). Thus, if \eqref{eq: ML-ortho-quantum speed limit} is saturated, so is \eqref{eq: MT-ortho-quantum speed limit}, and the reverse holds if the support of the initial state is not perpendicular to the eigenspace belonging to the smallest eigenvalue. These observations led Levitin and Toffoli to conclude that the maximum of $\tauMT(0)$ and $\tauML(0)$ is a tight lower bound for the evolution time only reachable for states such that $\langle H-\epsilon_0\rangle = \Delta H$. We can draw the same conclusion from the discussion in Sec.\ \ref{sec: The generalized Margolus-Levitin quantum speed limit}.

\begin{figure}[t]
	\centering
	\includegraphics[width=0.97\linewidth]{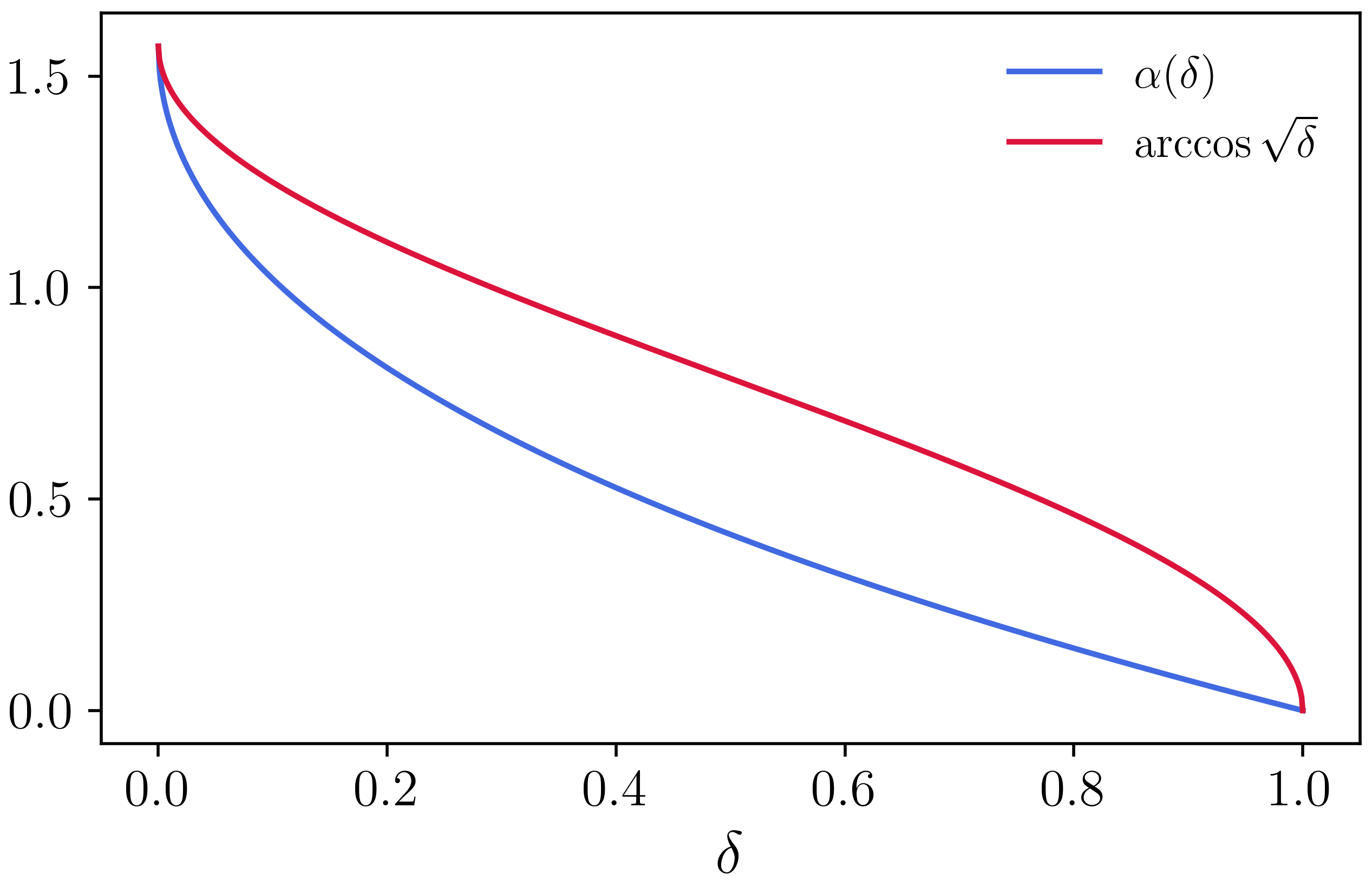}
	\caption{The graphs of $\alpha(\delta)$ and $\arccos\sqrt{\delta}$. The two functions agree only for $\delta=0$ and $\delta=1$. For all other fidelities, $\arccos\sqrt{\delta}$ is strictly greater than $\alpha(\delta)$.}
	\label{fig: alpha and arccos}
\end{figure}

Interestingly, the situation is almost the reverse for a nonzero fidelity $\delta$ between initial and final states; the Mandelstam–Tamm and the extended Margolus–Levitin QSLs can never saturate simultaneously: If the state follows a geodesic, then \eqref{eq: the Margolus-Levitin quantum speed limit} is not saturated by Sec.\ \ref{sec: 2D}, and if \eqref{eq: the Margolus-Levitin quantum speed limit} is an identity, the state does not follow a geodesic. However, the maximum QSL $\max\{\tauMT(\delta),\tauML(\delta)\}$ can be reached: If the state follows a geodesic, $\tau = \tauMT(\delta) > \tauML(\delta)$, and if condition \eqref{zeta} is satisfied, $\tau = \tauML(\delta) > \tauMT(\delta)$. In the latter case, $\langle H-\epsilon_0\rangle$ must differ from $\Delta H$ since $\alpha(\delta)$ is strictly less than $\arccos\sqrt{\delta}$ for $0<\delta<1$; see Fig.\ \ref{fig: alpha and arccos}.

\begin{rmk}
    The state does not follow a geodesic in a system that saturates the extended Margolus-Levitin QSL for a nonzero fidelity. However, the state follows a sub-Riemannian geodesic in a geodesic sphere centered at a ground state. See Appendix \ref{appendix B} for details.
\end{rmk}

\begin{rmk}
In quite a few papers it is claimed that $\arccos\sqrt{\delta}/\langle H-\epsilon_0\rangle$ is a QSL for isolated systems. Figure \ref{fig: alpha and arccos} and the fact that the extended Margolus-Levitin QSL is tight show that this is not true in general.
\end{rmk}

\subsection{The dual extended Margolus-Levitin quantum speed limit}\label{sec: the dual}
Let $\rho_a$ and $\rho_b$ be states with fidelity $\delta$, and suppose that $H$ evolves $\rho_a$ to $\rho_b$ in time $\tau$. Then $-H$ evolves $\rho_b$ to $\rho_a$ in time $\tau$. The smallest eigenvalue of $-H$ is $-\emax$, where $\emax$ is the largest eigenvalue of $H$, and according to the extended Margolus-Levitin QSL, 
\begin{equation}\label{eq: dual MLQSL}
    \tau\geq \tauML^*(\delta),\quad \tauML^*(\delta)=\frac{\alpha(\delta)}{\langle \emax - H \rangle}.
\end{equation}
This estimate generalizes the main result in Ref.\ \cite{NeAlSa2022} to an arbitrary fidelity between initial and final states. We adhere to the terminology in Ref.\ \cite{NeAlSa2022} and call $\tauML^*(\delta)$ the dual extended Margolus-Levitin QSL. Note that for some states, $\tauML^*(\delta)$ is greater than $\tauML(\delta)$ and $\tauMT(\delta)$. 

Since the estimate in \eqref{eq: dual MLQSL} is a consequence of applying the extended Margolus-Levitin QSL to evolution generated by $-H$, it follows from Appendix \ref{appendix A} that \eqref{eq: dual MLQSL} can be saturated in all dimensions and that, when so, the state is an effective qubit for $-H$ whose support is not orthogonal to the eigenspace corresponding to $-\emax$. But then the state is also an effective qubit for $H$ whose support is not orthogonal to the eigenspace corresponding to $\emax$. We call such an effective qubit \emph{partly maximally excited}. To summarize,
%
if $\tau \langle \emax-H \rangle$ assumes its smallest possible value when evaluated for all Hamiltonians $H$, states $\rho$, and $\tau\geq 0$ such that $H$ transforms $\rho$ into a state with fidelity $\delta$ in time $\tau$, then $\rho$ is a partly maximally excited effective qubit for $H$ and $ \tau \langle \emax-H \rangle = \alpha(\delta)$.

\begin{rmk}
If we restrict the set of states as in Remark \ref{rmk: ett}, $\epsilon_{\mathrm{max}}$ can be replaced by the largest occupied energy.
\end{rmk}

\begin{figure}[t]
	\centering
	\includegraphics[width=0.7\linewidth]{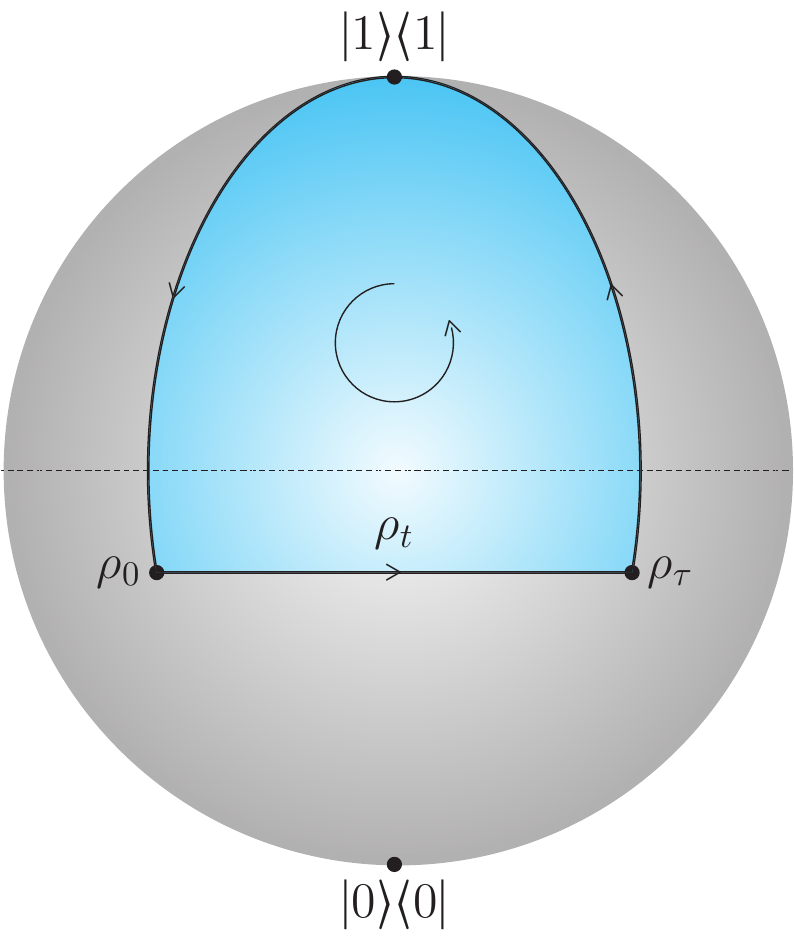}
	\caption{A Seifert surface in the Bloch sphere for the $\ketbra{1}{1}$ closure of an evolving qubit. The symplectic area of the surface is positive and equal to $\tau\langle\epsilon_{\mathrm{max}}-H\rangle$. Note that the surface is, in a sense, dual to the Seifert surface in Fig.\ \ref{fig: Symplectic area in the Bloch sphere}. If the dual extended Margolus-Levitin QSL is saturated, the surface assumes its smallest possible symplectic area. For not fully distinguishable initial and final states, the evolving state then has a strictly positive $z$ coordinate. This is in contrast to the case when the extended Margolus-Levitin QSL is saturated, in which case the $z$ coordinate is strictly negative. For fully distinguishable initial and final states, both QSLs can be saturated simultaneously. The magnitudes of the symplectic areas of the corresponding Seifert surfaces are then equal.}
	\label{fig: double Seifert}
\end{figure}

The discussion in Sec.\ \ref{sec: Geometry of the Margolus-Levitin quantum speed limit}, where we deliberately did not specify the eigenvalue $\epsilon$, tells us that
\begin{equation}\label{eq: furtitre}
    \tau\langle \emax-H \rangle 
    = -\int_{\rho_t} A^\sigma 
    = \iint_\Sigma\omega,
\end{equation}
where $\sigma$ is an eigenstate of $H$ with eigenvalue $\emax$ that has nonzero fidelity with the initial state. The surface $\Sigma$ is an arbitrary Seifert surface for the $\sigma$ closure $\rho_t^\sigma$ in $\Omega(\sigma)$. In Fig.\ \ref{fig: double Seifert} we have illustrated such a Seifert surface in the Bloch sphere for the same evolution as in Fig.\ \ref{fig: Symplectic area in the Bloch sphere}.
In this case, a concatenation of the evolution curve with the shortest geodesic connecting the initial and final states to the excited state $\ketbra{1}{1}$ parametrizes the boundary of the Seifert surface. Furthermore, the orientation of the boundary is such that the Seifert surface has a positive symplectic area, which is consistent with equation \eqref{eq: furtitre}. The $z$ coordinate of a qubit state that saturates the dual extended Margolus-Levitin QSL satisfies the equation
\begin{equation}
	\frac{1-z}{2}\arccos\bigg(\frac{2\delta-1-z^2}{1-z^2}\bigg)=\alpha(\delta).
\end{equation}
For $\delta\ne 0$, the $z$ coordinate of a saturating evolution is strictly positive in contrast to an evolution that saturates the ``original'' extended Margolus-Levitin QSL, in which case the $z$ coordinate is strictly negative, as in Fig.\ \ref{fig: Symplectic area in the Bloch sphere}. This observation lets us conclude that for not fully distinguishable initial and final states, the extended Margolus-Levitin QSL and its dual can never saturate simultaneously. Because if that were the case, the state would be an effective qubit with support in the span of a ground state and a highest energy state. In the projectivization of the span of these eigenstates, the $z$ coordinate of the Bloch vector of the system's state would be strictly positive and strictly negative, which is contradictory. For $\delta=0$, however, the QSLs can saturate simultaneously. The evolving state then follows a geodesic and
\begin{equation}
   \langle \emax-H \rangle = \Delta H= \langle H-\epsilon_0\rangle.
\end{equation}

\subsection{Approximations of the extended Margolus-Levitin quantum speed limit}
\label{sec: Approximations}
Consider a quantum system in a state $\rho$ with Hamiltonian $H$. Suppose the system evolves into a state with fidelity $\delta$ relative to $\rho$ in time $\tau$. Then 
\begin{equation}\label{eq: approx 1}
	\tau \geq \tau_1(\delta),\quad \tau_1(\delta)=\frac{1-\sqrt{\delta}}{\beta\langle H-\epsilon_0\rangle},
\end{equation}
where the requirement that $y=1-\beta x$ is a tangent line to $\cos x$ specifies $\beta$ ($\beta\approx 0.724$). Also,
\begin{align}
	&\tau \geq \tau_2(\delta),\quad \tau_2(\delta)=\frac{4\arccos^2\!\sqrt{\delta}}{\beta\pi^2\langle H-\epsilon_0\rangle}, \label{eq: approx 2} \\
	&\tau \geq \tau_3(\delta),\quad \tau_3(\delta)=\frac{2\arccos^2\!\sqrt{\delta}}{\pi\langle H-\epsilon_0\rangle}.\label{eq: approx 3}
\end{align}
Derivations of \eqref{eq: approx 1} and \eqref{eq: approx 2} are found in Refs.\ \cite{AnHe2014,An2018}, and \eqref{eq: approx 3} is mentioned in Ref.\ \cite{GiLlMa2003}. Where this latter QSL comes from is still unclear to the authors.

\begin{figure}[t]
	\centering
	\includegraphics[width=0.97\linewidth]{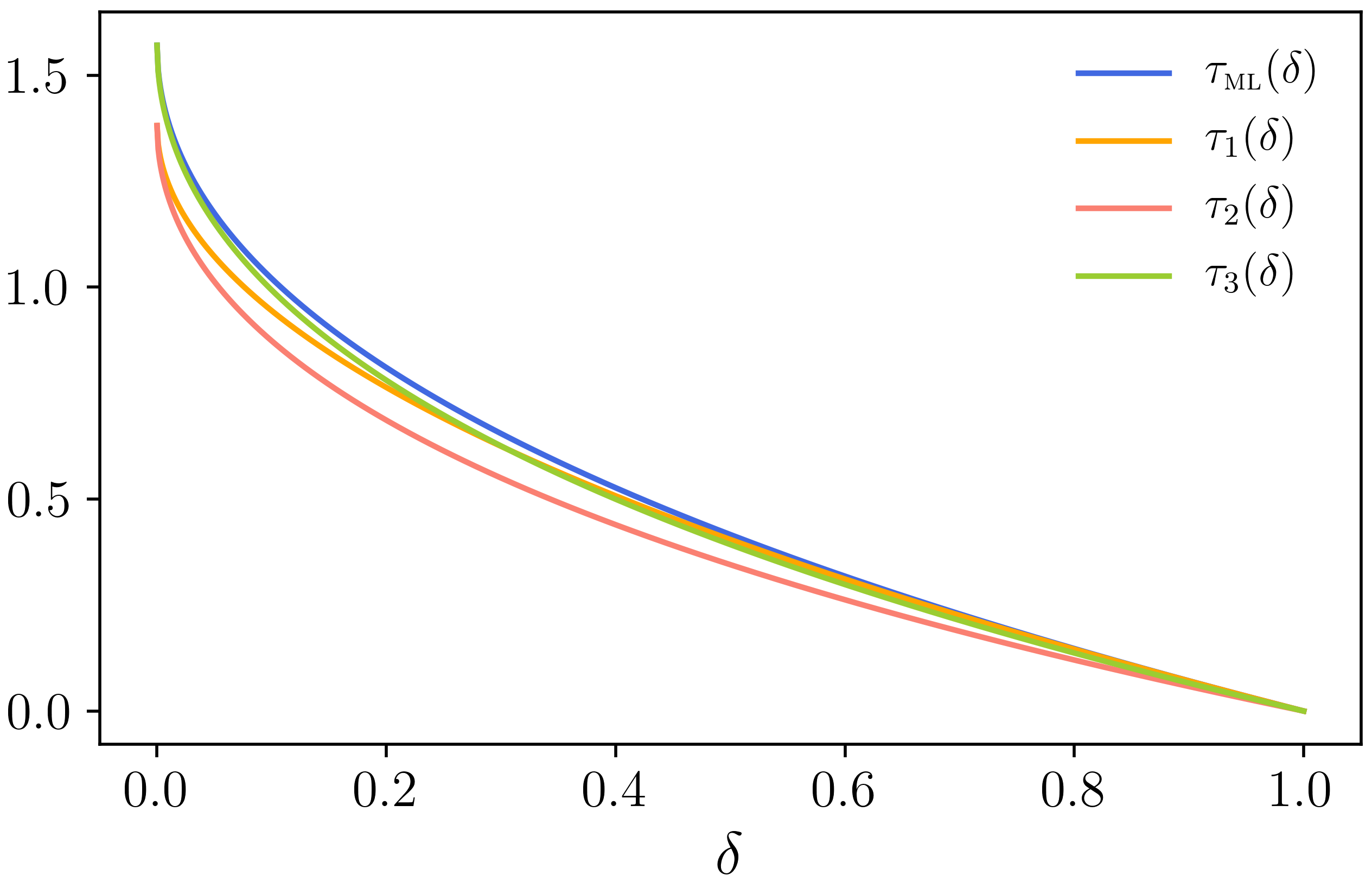}
	\caption{The graphs of $\tauML(\delta)$, $\tau_1(\delta)$, $\tau_2(\delta)$, and $\tau_3(\delta)$ multiplied by the normalized expected energy $\langle H-\epsilon_0\rangle$.
 The graph belonging to $\tauML(\delta)$ lies above the other graphs. Thus $\tauML(\delta)$ is the strongest bound. The weakest bound is $\tau_2(\delta)$. Since the graphs of $\tau_1(\delta)$ and $\tau_3(\delta)$ intersect, their mutual strength depends on the fidelity.}
	\label{fig: approximate quantum speed limits}
\end{figure}

Figure \ref{fig: approximate quantum speed limits} displays the graphs of the extended Margolus-Levitin QSL $\tauML(\delta)$ as well as the QSLs $\tau_1(\delta)$, $\tau_2(\delta)$, and $\tau_3(\delta)$ multiplied by $\langle H-\epsilon_0\rangle$. 
Apparently, $\tau_1(\delta)$, $\tau_2(\delta)$, and $\tau_3(\delta)$ are weaker than $\tauML(\delta)$, and $\tau_2(\delta)$ is weaker than $\tau_1(\delta)$ and $\tau_3(\delta)$. However, which of $\tau_1(\delta)$ and $\tau_3(\delta)$ is the stronger QSL depends on the fidelity $\delta$, with $\tau_1(\delta)$ being greater than $\tau_3(\delta)$ for large values of $\delta$ and $\tau_3(\delta)$ being greater than $\tau_1(\delta)$ for small values of $\delta$.

\section{Summary and outlook}\label{sec: Outlook}
Giovannetti \emph{et al.}\ \cite{GiLlMa2003} showed, partly numerically, that if an isolated system evolves between two states with fidelity $\delta$, then the evolution time is bounded from below by the extended Margolus-Levitin QSL \eqref{eq: the Margolus-Levitin quantum speed limit}. Giovannetti \emph{et al.}\ also showed that this QSL is tight in all dimensions. 

In this paper, we have derived the extended Margolus-Levitin QSL analytically and characterized the systems for which this QSL is saturated. Furthermore, we have interpreted the extended Margolus-Levitin QSL geometrically as an extremal dynamical phase in a gauge specified by the system’s ground state. We have also shown that the maximum of the Mandelstam-Tamm and the extended Margolus-Levitin QSLs is a tight QSL that behaves differently depending on whether or not the initial state and the final state are fully distinguishable. In addition, we have derived a tight dual version of the extended Margolus-Levitin QSL using a straightforward time-reversal argument. The dual QSL has similar properties as the extended Margolus-Levitin QSL and saturates under similar circumstances but involves the largest rather than the smallest occupied energy. We showed that the two QSLs can only be saturated simultaneously if the start and end states are fully distinguishable. We concluded the paper by reproducing three QSLs related to, but slightly weaker than, the extended Margolus-Levitin QSL. 

A recent paper on evolution time estimates for closed systems \cite{HoSo2023} suggests that the Margolus-Levitin QSL does not straightforwardly extend to systems whose dynamics are governed by time-dependent Hamiltonians, at least not without limitations on the width of the energy spectrum.

The geometric analysis performed here shows that extended Margolus-Levitin QSL is closely related to the Aharonov-Anandan geometric phase \cite{AhAn1987}. This observation is further elaborated in Ref.\ \cite{HoSo2023c} where QSLs for cyclically evolving systems are derived.

Mixed state QSLs resembling the Margolus-Levitin QSL exist \cite{AnAl2004, An2018, AnHe2014}, and Giovannetti \emph{et al.}\ \cite{GiLlMa2003} showed that also the extended Margolus-Levitin QSL can be generalized to a QSL for systems in mixed states, with $\delta$ being the Uhlmann fidelity between the initial and final states \cite{Uh1976}. Whether this generalization has a symplectic-geometric interpretation similar to the one presented here is an open question. So is the question whether the generalized extended Margolus-Levitin QSL connects to a geometric phase for mixed states \cite{An2018, AnHe2015, AnHe2013, SjPaEkAnErOiVe2000, Uh1995}. The authors intend to investigate these questions in a forthcoming paper.

\section*{Acknowledgments}\label{sec: Acknowledgments}
We thank Dan Allan for valuable discussions.

\onecolumngrid
\appendix
\section{The extended Margolus-Levitin quantum speed limit for qubit systems}\label{appendix 0}
Let $H$ be the Hamiltonian in \eqref{eq: qubit Hamiltonian} and $\rho$ be a qubit state whose Bloch vector $\mathbf{r}=(x,y,z)$ is defined by \eqref{x}--\eqref{z}.
Then
\begin{equation}
\bra{0}\rho\ket{0} = \frac{1}{2}(1-z),\quad
\bra{1}\rho\ket{1}= \frac{1}{2}(1+z),\quad
\bra{0}\rho\ket{1}= \frac{1}{2}(x+iy),\quad
\bra{1}\rho\ket{0} = \frac{1}{2}(x-iy).
\end{equation}
Equation \eqref{eq: the z-coordinate} follows from
\begin{equation}
    2\langle H\rangle
    =2\bra{0}H\rho\ket{0}+2\bra{1}H\rho\ket{1}
    =2\epsilon_0\bra{0}\rho\ket{0}+2\epsilon_1\bra{1}\rho\ket{1}
    =\epsilon_0(1-z)+\epsilon_1(1+z)
    =\epsilon_1+\epsilon_0+(\epsilon_1-\epsilon_0)z.
\end{equation}

To prove equation \eqref{eq: inner product and fidelity} consider two qubit states $\rho_a$ and $\rho_b$ with Bloch vectors $\mathbf{r}_a=(x_a,y_a,z_a)$ and $\mathbf{r}_b=(x_b,y_b,z_b)$, respectively. Let $\delta$ be the fidelity between $\rho_a$ and $\rho_b$. Then
\begin{equation}
\begin{split}
    \delta    &=\bra{0}\rho_a\rho_b\ket{0}+\bra{1}\rho_a\rho_b\ket{1}\\    &=\bra{0}\rho_a\ket{0}\bra{0}\rho_b\ket{0}
    +\bra{0}\rho_a\ket{1}\bra{1}\rho_b\ket{0}+\bra{1}\rho_a\ket{0}\bra{0}\rho_b\ket{1}+\bra{1}\rho_a\ket{1}\bra{1}\rho_b\ket{1}\\
    &=\frac{1}{4}\big((1-z_a)(1-z_b)+(x_a+iy_a)(x_b-iy_b)+(x_a-iy_a)(x_b+iy_b)+(1+z_a)(1+z_b)\big)\\
    &=\frac{1}{2}(1+x_ax_b+y_ay_b+z_az_b)\\
    &=\frac{1}{2}(1+\mathbf{r}_a\cdot\mathbf{r}_b).
\end{split}
\end{equation}
This proves equation \eqref{eq: inner product and fidelity}. 

The dynamics caused by $H$ can be described as a rigid rotation of the Bloch sphere about the $z$ axis. To prove \eqref{tolv} assume $z_a=z_b=z$ and that $\mathbf{r}_a$ rotates to $\mathbf{r}_b$ during the evolution. Since the $z$ coordinate remains fixed, $\mathbf{r}_a$ follows a path of the same length as the orthogonal projection $\bar{\mathbf{r}}_a$ of $\mathbf{r}_a$ on the $xy$ plane. The projection $\bar{\mathbf{r}}_a$ rotates to the projection $\bar{\mathbf{r}}_b$ of $\mathbf{r}_b$ along a circular arc in the $xy$ plane. The radius of the corresponding circle sector is
\begin{equation}
    \sqrt{\bar{\mathbf{r}}_a\cdot \bar{\mathbf{r}}_a}=\sqrt{\mathbf{r}_a\cdot \mathbf{r}_a-z^2}=\sqrt{1-z^2}.
\end{equation} 
Furthermore, according to \eqref{eq: inner product and fidelity}, the circle sector has apex angle 
\begin{equation}
    \arccos\bigg(\frac{\bar{\mathbf{r}}_a\cdot \bar{\mathbf{r}}_b}{\bar{\mathbf{r}}_a\cdot \bar{\mathbf{r}}_a}\bigg)
    =\arccos\bigg(\frac{\mathbf{r}_a\cdot \mathbf{r}_b-z^2}{\mathbf{r}_a\cdot \mathbf{r}_a-z^2}\bigg)=\arccos\bigg(\frac{2\delta-1-z^2}{1-z^2}\bigg).
\end{equation}
Thus, the length of the circular arc is
\begin{equation}
    \sqrt{1-z^2}\arccos\bigg(\frac{2\delta-1-z^2}{1-z^2}\bigg).
\end{equation}

Finally, to prove \eqref{tretton} we first determine the speed of the rotating Bloch vector. Write $x_t$, $y_t$, and $z_t=z$ for the coordinates of the rotating Bloch vector. We have 
\begin{equation}
     \dot x_t+i\dot y_t=2\bra{0}\dot\rho_t\ket{1}=-2i\bra{0}[H,\rho_t]\ket{1}=-i(\epsilon_0-\epsilon_1)(x_t+iy_t).
\end{equation}
Hence $\dot x_t=(\epsilon_0-\epsilon_1)y_t$ and $\dot y_t=(\epsilon_1-\epsilon_0)x_t$. The rotating Bloch vector thus has the speed
\begin{equation}
    \sqrt{\dot x_t^2+\dot y_t^2}=(\epsilon_1-\epsilon_0)\sqrt{x_t^2+y_t^2}=(\epsilon_1-\epsilon_0)\sqrt{1-z^2}.
\end{equation}
This equation also says that $H$ causes the Bloch sphere to rotate with the angular speed $\epsilon_1-\epsilon_0$ about the $z$ axis. The evolution time equals the distance traveled by the Bloch vector divided by the speed of the Bloch vector,
\begin{equation}
    \tau=\frac{\sqrt{1-z^2}\arccos\big(\frac{2\delta-1-z^2}{1-z^2}\big)}{(\epsilon_1-\epsilon_0)\sqrt{1-z^2}}
    =\frac{1}{\epsilon_1-\epsilon_0}
    \arccos\bigg(\frac{2\delta-1-z^2}{1-z^2}\bigg).
\end{equation}
This is the statement in equation \eqref{tretton}.

\section{Characterization of time-optimal systems}\label{appendix A}
Consider a quantum system of dimension $n\geq 2$. Let $0\leq\delta <1$ and set
\begin{equation}\label{genalpha}
    \alpha(n,\delta)=\inf_{(H,\rho,\tau)}\tau\langle H-\epsilon_0\rangle,
\end{equation}
where the infimum is over all triples consisting of a Hamiltonian $H$, a state $\rho$, and a time $\tau\geq 0$ such that 
\begin{equation}\label{transform}
    \tr(\rho e^{-i\tau H}\rho e^{i\tau H})=\delta.
\end{equation}
As before, $\epsilon_0$ is the smallest eigenvalue of $H$. In this appendix we show that some triple $(H,\rho,\tau)$ realizes the infimum $\alpha(n,\delta)$ and that for any such triple, $\rho$ is a partly grounded effective qubit for $H$. Furthermore, we show that $\alpha(n,\delta)$ is the same for all $n$ and thus equals $\alpha(\delta)$ defined in equation \eqref{alpha}. Note that condition \eqref{transform} says that if $H$ controls the dynamics of the system, then $\rho$ will evolve to a state with fidelity $\delta$ in time $\tau$.

\subsection{Reduction to admissible pairs}
We begin by establishing that
\begin{equation}\label{normed genalpha}
	\alpha(n,\delta) = \inf_{(H,\rho)}\langle H\rangle,
\end{equation}
where the infimum is over all pairs consisting of a Hamiltonian $H$ with spectrum in $[0, 2\pi]$ and a state $\rho$ such that \begin{equation}\label{normed transform}
    \tr(\rho e^{-i H}\rho e^{i H})=\delta.
\end{equation}
We call such a pair \emph{admissible}.

Take an arbitrary triple $(H,\rho,\tau)$ that satisfies condition \eqref{transform}. Let $\rho'=e^{-i\tau H}\rho e^{i\tau H}$ and set $H_1=H-\epsilon_0$. The Hamiltonian $H_1$ is positive, has the minimum eigenvalue $0$, and gives rise to dynamics identical to that caused by $H$. Hence, $(H_1,\rho,\tau)$ satisfies condition \eqref{transform}. Since $H_1$ and $H$ have the same eigenspaces and, in particular, they have the same eigenspace for their smallest eigenvalues, $\rho$ is a partly grounded effective qubit for $H_1$ if and only if $\rho$ is a partly grounded effective qubit for $H$. 

Next, we normalize the evolution time by setting $H_2=\tau H_1$. The Hamiltonian $H_2$ is positive, has the minimum eigenvalue $0$, and evolves $\rho$ to $\rho'$ along the same path as $H_1$ but in time $1$. Thus $(H_2,\rho,1)$ satisfies condition \eqref{transform}. Since $H_2$ and $H_1$ have the same eigenspaces, especially for the eigenvalue $0$, $\rho$ is a partly grounded effective qubit for $H_2$ if and only if $\rho$ is a partly grounded effective qubit for $H_1$. 

Lastly, we define $H_3$ as the Hamiltonian obtained from $H_2$ by replacing each eigenvalue $\epsilon$ of $H_2$ with the $\epsilon'$ in $[0,2\pi]$ satisfying $\epsilon'=\epsilon\;\mathrm{mod}\,2\pi$. The Hamiltonian $H_3$ is positive, has the smallest eigenvalue $0$, and evolves $\rho$ to $\rho'$ in time $1$. The expectation value of $H_3$ is not greater than that of $H_2$. We thus have that
\begin{equation}\label{intressant}
	\langle H_3\rangle 
	\leq \langle H_2 \rangle 
	= \tau \langle H_1 \rangle 
	= \tau \langle H - \epsilon_0 \rangle.
\end{equation}
This shows that $\inf_{(H,\rho)}\langle H\rangle \leq \alpha(n,\delta)$, where the infimum is over all admissible pairs. The reverse inequality is automatically satisfied. Since the space of admissible pairs is compact, $\alpha(n,\delta)$ is attained for an admissible pair.

Now, suppose that $\rho$ is a partly grounded effective qubit for $H$ and thus for $H_2$. Then $\rho$ is also a partly grounded effective qubit for $H_3$. For eigenspaces of $H_2$ can merge when forming $H_3$, but not in such a way that the support of $\rho$ is covered by the eigenspace of $H_3$ that belongs to the eigenvalue $0$. The reverse, however, is not necessarily true. If $\rho$ is a partly grounded effective qubit for $H_3$, $\rho$ need not be a partly grounded effective qubit for $H_2$. But in that case, the inequality in \eqref{intressant} is strict, and the original triple $(H,\rho,\tau)$ does not saturate \eqref{genalpha}.

\begin{rmk}\label{rmk: sju}
One can choose the state arbitrarily and further restrict the infimum to Hamiltonians that form an admissible pair with the selected state. For if $(H,\rho,\tau)$ satisfies \eqref{transform}, $\rho'$ is an arbitrary state, $U$ is a unitary such that $U\rho U^\dagger=\rho'$, and $H'=UHU^\dagger$, then also $(H',\rho',\tau)$ satisfies \eqref{transform}. Furthermore, $H'$ and $H$ have the same spectrum, the expectation value of $H'$ when the state is $\rho'$ is the same as the expectation value of $H$ when the state is $\rho$, and $\rho'$ is a partly grounded effective qubit for $H'$ if and only if $\rho$ is a partly grounded effective qubit for $H$.
\end{rmk}

\subsection{Not fully distinguishable initial and final states}\label{not dist}
Assume that $\delta>0$. (For the sake of completeness, we treat the case $\delta=0$ in Appendix \ref{full dist}, although this case has been dealt with earlier \cite{SoBjTsTr1999, LeTo2009}.) Let $(\mathbf{p},\boldsymbol{\epsilon})$ denote a point $(p_1,\dots,p_n,\epsilon_1,\dots,\epsilon_n)$ in the compact rectangular block $[0,1]^n\times [0,2\pi]^n$. Define the functions $f$, $g$, and $h$ on $[0,1]^n\times [0,2\pi]^n$ as
\begin{align}
    &f(\mathbf{p},\boldsymbol{\epsilon})=\sum_{j=1}^{n} p_j\epsilon_j, \\
    &g(\mathbf{p},\boldsymbol{\epsilon})=\Big(\sum_{j=1}^{n} p_j\cos\epsilon_j\Big)^2+\Big(\sum_{j=1}^{n} p_j\sin\epsilon_j\Big)^2, \\
    &h(\mathbf{p},\boldsymbol{\epsilon})=\sum_{j=1}^{n} p_j.
\end{align}
Furthermore, let $M$ be the subset of $[0,1]^n\times [0,2\pi]^n$ defined by the constraints $g(\mathbf{p},\boldsymbol{\epsilon})=\delta$ and $h(\mathbf{p},\boldsymbol{\epsilon})=1$. Since $M$ is compact and $f$ is continuous, the restriction of $f$ to $M$, denoted $f|_M$, assumes a minimum value. This minimum value is $\alpha(n,\delta)$. To see this suppose that $(H,\rho)$ is an admissible pair such that $\langle H\rangle =\alpha(n,\delta)$. Let $\epsilon_1,\epsilon_2,\dots,\epsilon_n$ be the eigenvalues of $H$, with corresponding eigenvectors $\ket{1},\dots,\ket{n}$, and set $p_j=\bra{j}\rho\ket{j}$. The point $(\mathbf{p},\boldsymbol{\epsilon})=(p_1,\dots,p_n,\epsilon_1,\dots,\epsilon_n)$ belongs to $M$, since $g(\mathbf{p},\boldsymbol{\epsilon})=\tr(\rho e^{-iH}\rho e^{iH})=\delta$ and $h(\mathbf{p},\boldsymbol{\epsilon})=\tr (\rho)=1$, and $f(\mathbf{p},\boldsymbol{\epsilon})=\langle H\rangle =\alpha(n,\delta)$. Thus, $f|_M$ assumes the value $\alpha(n,\delta)$. To see that $\alpha(n,\delta)$ is the smallest value of $f|_M$ let $(\mathbf{p},\boldsymbol{\epsilon})$ is an arbitrary point in $M$. Then, let $\ket{1},\dots,\ket{n}$ be an arbitrary orthonormal basis for the Hilbert space of the system and define $H=\sum_{j}\epsilon_j\ketbra{j}{j}$ and $\rho =\sum_{i,j}\sqrt{p_ip_j}\ketbra{i}{j}$. The pair $(H,\rho)$ is admissible and $f(\mathbf{p},\boldsymbol{\epsilon})=\langle H\rangle$. Hence, $f(\mathbf{p},\boldsymbol{\epsilon})\geq\alpha(n,\delta)$.

Let $(\mathbf{p},\boldsymbol{\epsilon})$ be a point in $M$ at which $f|_M$ assumes its minimum value and let $J$ be the set of indices $j$ for which $p_j\ne 0$. The minimum point $(\mathbf{p},\boldsymbol{\epsilon})$ has the following properties:
\begin{itemize}
	\item[(i)] There are indices $j_1$ and $j_2$ in $J$ such that $\epsilon_{j_1}\ne 0$ and $\epsilon_{j_2}= 0$.
	\item[(ii)] For all indices $j_1$ and $j_2$ in $J$ such that $\epsilon_{j_1}\ne 0$ and $\epsilon_{j_2}\ne 0$ we have that $\epsilon_{j_1} = \epsilon_{j_2}$.
\end{itemize} 
Before showing that $(\mathbf{p},\boldsymbol{\epsilon})$ has these properties let us discuss their implications for the extended Margolus-Levitin QSL.

Assume that $(H,\rho)$ is an admissible pair such that $\langle H\rangle = \alpha(n,\delta)$. Let $\epsilon_1,\epsilon_2,\dots,\epsilon_{n}$ be the spectrum of $H$ and let $p_j$ be the probability of obtaining $\epsilon_j$ when measuring $H$ in the state $\rho$ divided by the degeneracy of $\epsilon_j$. The point $(\mathbf{p},\boldsymbol{\epsilon})=(p_1,\dots,p_{n},\epsilon_1,\dots,\epsilon_{n})$ belongs to $M$ since $g(\mathbf{p},\boldsymbol{\epsilon})=\delta$ according to the normed evolution time condition  \eqref{normed transform} and $h(\mathbf{p},\boldsymbol{\epsilon})=1$ according to the law of total probability. Furthermore, $f|_M$ assumes its minimum value at $(\mathbf{p},\boldsymbol{\epsilon})$ since $f(\mathbf{p},\boldsymbol{\epsilon})=\langle H\rangle$. Property (i) says that $\rho$ is partly grounded; that is, its support is not orthogonal to the eigenspace belonging to the smallest eigenvalue of $H$, which in this case is $0$. Property (ii) says that $\rho$ is an effective qubit because the support of $\rho$ is only non-orthogonal to one eigenspace other than that belonging to $0$.

Let us now prove properties (i) and (ii). For clarity let $(\mathbf{p},\boldsymbol{\epsilon})=(p_1,\dots,p_{n},\epsilon_1,\dots,\epsilon_{n})$ denote an arbitrary, unspecified point in $[0,1]^n\times [0,2\pi]^n$ and let $(\mathbf{p}^*,\boldsymbol{\epsilon}^*)=(p_1^*,\dots,p_{n}^*,\epsilon_1^*,\dots,\epsilon_{n}^*)$ be a point in $M$ at which $f|_M$ assumes its minimum value. Let $k$ be the number of nonzero $p_j^*$s. Since the values of $f$, $g$, and $h$ are invariant under rearrangements of the form $(p_1,\dots,p_{n},\epsilon_1,\dots,\epsilon_{n})\to (p_{\kappa(1)},\dots,p_{\kappa(n)},\epsilon_{\kappa(1)},\dots,\epsilon_{\kappa(n)})$, where $\kappa$ is a permutation of the set $\{1,2,\dots,n\}$, we can assume that the $k$ nonzero $p_j^*$s are placed first in the sequence $\mathbf{p}^*$, so that $\mathbf{p}^*=(p_1^*,\dots,p_{k}^*,\bar{\mathbf{0}}^*)$, where $\bar{\mathbf{0}}^*$ is the vector consisting of $n-k$ zeros, and that the corresponding $\epsilon_j^*$s are arranged in descending order of magnitude. We assume that the number of nonzero $\epsilon_j^*$s among the first $k$ is $l$. These constitute the first $l$ elements of $\boldsymbol{\epsilon}^*$. We denote the remaining $n-l$ $\epsilon_j^*$s by $\bar{\boldsymbol{\epsilon}}^*$, so that $\boldsymbol{\epsilon}^*=(\epsilon_1^*,\dots,\epsilon_{l}^*,\bar{\boldsymbol{\epsilon}}^*)$. The minimum point has the following properties:
\begin{itemize}
\item[(a)] $\epsilon_1^*,\dots,\epsilon_l^*$ are strictly less than $2\pi$. For suppose that some $\epsilon_j^*=2\pi$. If we change the value of this $\epsilon_j^*$ to $0$, $g$ and $h$ will not change their values, and hence we are still in $M$. However, since $p_j\ne 0$, such a change lowers the value of $f$. This contradicts that $f|_M$ assumes its minimum value at $(\mathbf{p}^*,\boldsymbol{\epsilon}^*)$.
\item[(b)] There exists a $j$ such that $p_j^*\ne 0$ and $\epsilon_j^*=0$, and thus $k>l$. For suppose that $k=l$. The functions $g$ and $h$ assume the same values at $(\mathbf{p}^*,\boldsymbol{\epsilon}^*)$ and $(\mathbf{p}^*,\epsilon_1^*-\epsilon_k^*,\dots,\epsilon_k^*-\epsilon_k^*,0,\dots,0)$, so both points belong to $M$. However, since $p^*_k\ne 0$, $f$ has a lower value at the latter point. This contradicts that $f|_M$ assumes its minimum value in $(\mathbf{p}^*,\boldsymbol{\epsilon}^*)$. Since not all the $\epsilon_j^*$s can be zero, statement (i) follows.
\end{itemize}

It remains to prove statement (ii), that is, that all the nonzero $\epsilon_1^*,\dots,\epsilon_l^*$ have the same value. Since we have assumed that $\delta>0$, there exists a unique function $\theta$ on $M$ with values in the interval $(-\pi,\pi]$ such that
\begin{align}
    &\sqrt{\delta} \cos\theta(\mathbf{p},\boldsymbol{\epsilon}) = \sum_{j=1}^n p_j\cos\epsilon_j, \\
    &\sqrt{\delta} \sin\theta(\mathbf{p},\boldsymbol{\epsilon}) = \sum_{j=1}^n p_j\sin\epsilon_j.
\end{align}
This follows immediately from the observation that $g(\mathbf{p},\boldsymbol{\epsilon})=\delta$ if and only if $\sum_j p_j e^{i\epsilon_j}=\sqrt{\delta}e^{i\theta(\mathbf{p},\boldsymbol{\epsilon})}$ for a unique $\theta(\mathbf{p},\boldsymbol{\epsilon})$ in $(-\pi,\pi]$.
Now, consider the restrictions of $f$, $g$, and $h$ to $F=[0,1]^k \times \{\bar{\mathbf{0}}^*\} \times [0,2\pi]^l\times\{\bar{\boldsymbol{\epsilon}}^*\}$:
\begin{align}
    &f|_F(\mathbf{p},\boldsymbol{\epsilon})=\sum_{j=1}^l p_j\epsilon_j, \\
    &g|_F(\mathbf{p},\boldsymbol{\epsilon})=\Big(\sum_{j=1}^l p_j\cos\epsilon_j + \sum_{j=l+1}^kp_j\Big)^2+\Big(\sum_{j=1}^l p_j\sin\epsilon_j\Big)^2, \label{seetolv}\\
    &h|_F(\mathbf{p},\boldsymbol{\epsilon})=\sum_{j=1}^k p_j.
\end{align}
The point $(\mathbf{p}^*,\boldsymbol{\epsilon}^*)$ lies in the intersection of $M$ and the interior of $F$. Let $\theta^*=\theta(\mathbf{p}^*,\boldsymbol{\epsilon}^*)$. The gradient vectors of $f|_F$, $g|_F$, and $h|_F$ at $(\mathbf{p}^*,\boldsymbol{\epsilon}^*)$ are 
\begin{align}
    \nabla f|_F(\mathbf{p}^*,\boldsymbol{\epsilon}^*) 
        &= \sum_{j=1}^l \epsilon_j^*\frac{\partial}{\partial p_j} + \sum_{j=1}^l p_j^* \frac{\partial}{\partial \epsilon_j}, \\
    \nabla g|_F(\mathbf{p}^*,\boldsymbol{\epsilon}^*)
        &= 2\sqrt{\delta}\Big(\sum_{j=1}^l\cos(\theta^*-\epsilon_j^*)\frac{\partial}{\partial p_j}+\sum_{j=l+1}^k\cos\theta^*
        \frac{\partial}{\partial p_j}+\sum_{j=1}^l
        p_j^* \sin(\theta^*-\epsilon_j^*)\frac{\partial}{\partial \epsilon_j}\Big), \label{seefemton}\\
    \nabla h|_F(\mathbf{p}^*,\boldsymbol{\epsilon}^*)
        &=\sum_{j=1}^k \frac{\partial}{\partial p_j}. 
\end{align}
None of these are zero. According to the method of Lagrange multipliers, there are real and nonzero constants $\lambda$ and $\mu$ such that $\nabla f|_F(\mathbf{p}^*,\boldsymbol{\epsilon}^*) = \lambda \nabla g|_F(\mathbf{p}^*,\boldsymbol{\epsilon}^*) +  \mu\nabla  h|_F(\mathbf{p}^*,\boldsymbol{\epsilon}^*)$. We thus have that
\begin{align}
    \epsilon_j^* &= 2\sqrt{\delta}\lambda\cos(\theta^*-\epsilon_j^*) + \mu, \label{seesjutton} \\
    1 &= 2\sqrt{\delta}\lambda \sin(\theta^*-\epsilon_j^*), \label{seearton} \\
    0 &= 2\sqrt{\delta}\lambda\cos\theta^* + \mu. \label{seesjuttonplushalv}
\end{align}
Equations \eqref{seesjutton} and \eqref{seearton} hold for $j=1,2,\dots,l$. These equations imply that all the $\epsilon_j^*$s are solutions to the quadratic equation $(x-\mu)^2=4\delta\lambda^2-1$. For if we move $\mu$ to the left side in \eqref{seesjutton}, square both \eqref{seesjutton} and \eqref{seearton}, and then add them, we obtain the equation $(\epsilon_j^*-\mu)^2=4\delta\lambda^2-1$. Thus, the $\epsilon_j^*$s can assume at most two values, 
\begin{align}
	&\bar\epsilon_1=\mu+\sqrt{4\delta\lambda^2-1},\label{sextiett}\\
	&\bar\epsilon_2=\mu-\sqrt{4\delta\lambda^2-1}.\label{sextitva}
\end{align}
We assume that both of these are present among the $\epsilon_j^*$s. (Otherwise, we are done.) According to \eqref{sextiett} and \eqref{sextitva}, $\mu$ is the arithmetic mean of $\bar\epsilon_1$ and $\bar\epsilon_2$. Furthermore, \eqref{seesjutton} and \eqref{seearton} imply that
\begin{align}
    2\cos\Big(\theta^*-\frac{\bar\epsilon_1+\bar\epsilon_1}{2}\Big)
    	&\cos\Big(\frac{\bar\epsilon_2-\bar\epsilon_1}{2}\Big)
	  = \cos(\theta^*-\bar\epsilon_1)+\cos(\theta^*-\bar\epsilon_2)
  	= 0, \\
    2\cos\Big(\theta^*-\frac{\bar\epsilon_1+\bar\epsilon_1}{2}\Big)
   	&\sin\Big(\frac{\bar\epsilon_2-\bar\epsilon_1}{2}\Big)
		= \sin(\theta^*-\bar\epsilon_1)-\sin(\theta^*-\bar\epsilon_2)
   = 0.
\end{align}
Thus,
\begin{equation}
	\theta^*=\frac{\bar\epsilon_1+\bar\epsilon_2}{2} + n\cdot\frac{\pi}{2}
\end{equation}
for an odd integer $n$. Consequently,
\begin{align}
    &\sin(\theta^*-\bar\epsilon_1)
    =(-1)^{\frac{n-1}{2}}\cos\left(\frac{\bar\epsilon_1-\bar\epsilon_2}{2}\right),\\
    &\cos(\theta^*-\bar\epsilon_1)
    =(-1)^{\frac{n-1}{2}}\sin\left(\frac{\bar\epsilon_1-\bar\epsilon_2}{2}\right).
\end{align}
Together with \eqref{seesjutton} och \eqref{seearton}, these equations imply
\begin{equation}
    \frac{\bar\epsilon_1-\bar\epsilon_2}{2}
    = \frac{2\sqrt{\delta}\lambda\cos(\theta^*-\bar\epsilon_1)}{2\sqrt{\delta}\lambda\sin(\theta^*-\bar\epsilon_1)}
    = \tan\left(\frac{\bar\epsilon_1-\bar\epsilon_2}{2}\right).
\end{equation}
Since $-\pi\leq (\bar\epsilon_1-\bar\epsilon_2)/2\leq\pi$ and $x=\tan x$ has the unique solution $x=0$ in this interval, we can conclude that $\bar\epsilon_1=\bar\epsilon_2$.

\subsection{Fully distinguishable initial and final states}\label{full dist}
If $\delta=0$ we must modify the previous section's arguments slightly. We start by replacing $g$ with the two functions
\begin{align}
    &g_1(\mathbf{p},\boldsymbol{\epsilon})=\sum_{j=1}^n p_j\cos\epsilon_j,\\
    &g_2(\mathbf{p},\boldsymbol{\epsilon})=\sum_{j=1}^n p_j\sin\epsilon_j,
\end{align}
and define $M$ as the subset of $[0,1]^n\times [0,2\pi]^n$ given by the constraints $g_1(\mathbf{p},\boldsymbol{\epsilon})=0$, $g_2(\mathbf{p},\boldsymbol{\epsilon})=0$, and $h(\mathbf{p},\boldsymbol{\epsilon})=1$. We again let $(\mathbf{p}^*,\boldsymbol{\epsilon}^*)=(p_1^*,\dots,p_n^*,\epsilon_1^*,\dots,\epsilon_n^*)$ be a point in $M$ at which $f|_M$ assumes its minimum value, which in this case is $\alpha(n,0)=\pi/2$. We arrange the coordinates in $(\mathbf{p}^*,\boldsymbol{\epsilon}^*)$ as in Appendix \ref{not dist}. Properties (a) and (b) of $(\mathbf{p}^*,\boldsymbol{\epsilon}^*)$ also apply in this case. We consider the restrictions of $f$, $g_1$, $g_2$, and $h$ to $F=[0,1]^k \times \{\bar{\mathbf{0}}^*\} \times [0,2\pi]^l\times\{\bar{\boldsymbol{\epsilon}}^*\}$:
\begin{align}
    &f|_F(\mathbf{p},\boldsymbol{\epsilon})=\sum_{j=1}^l p_j\epsilon_j, \\
    &g_1|_F(\mathbf{p},\boldsymbol{\epsilon})=\sum_{j=1}^l p_j\cos\epsilon_j + \sum_{j=l+1}^kp_j,\\
    &g_2|_F(\mathbf{p},\boldsymbol{\epsilon})=\sum_{j=1}^l p_j\sin\epsilon_j,\\
    &h|_F(\mathbf{p},\boldsymbol{\epsilon})=\sum_{j=1}^k p_j.
\end{align}
The point $(\mathbf{p}^*,\boldsymbol{\epsilon}^*)$ lies in the intersection of $M$ and the interior of $F$, and the gradient vectors of the restrictions of $f$, $g$, and $h$ to $F$ at $(\mathbf{p}^*,\boldsymbol{\epsilon}^*)$ are
\begin{align}
    \nabla f|_F(\mathbf{p}^*,\boldsymbol{\epsilon}^*) 
        &= \sum_{j=1}^l \epsilon_j^*\frac{\partial}{\partial p_j} + \sum_{j=1}^l p_j^* \frac{\partial}{\partial \epsilon_j}, \\
    \nabla g_1|_F(\mathbf{p}^*,\boldsymbol{\epsilon}^*) 
        &= \sum_{j=1}^l\cos\epsilon_j^*\frac{\partial}{\partial p_j}+\sum_{j=l+1}^k
        \frac{\partial}{\partial p_j}-\sum_{j=1}^l
        p_j^* \sin\epsilon_j^*\frac{\partial}{\partial \epsilon_j},\\
        \nabla g_2|_F(\mathbf{p}^*,\boldsymbol{\epsilon}^*) 
        &= \sum_{j=1}^l\sin\epsilon_j^*\frac{\partial}{\partial p_j}+\sum_{j=1}^l
        p_j^* \cos\epsilon_j^*\frac{\partial}{\partial \epsilon_j},\\
           \nabla h|_F(\mathbf{p}^*,\boldsymbol{\epsilon}^*)
        &=\sum_{j=1}^k \frac{\partial}{\partial p_j}.
\end{align}
None of these are zero. According to the method of Lagrange multipliers there exist real nonzero constants $\lambda_1$, $\lambda_2$, and $\mu$ such that $\nabla f|_F(\mathbf{p}^*,\boldsymbol{\epsilon}^*) = \lambda_1 \nabla g_1|_F(\mathbf{p}^*,\boldsymbol{\epsilon}^*) + \lambda_2 \nabla g_2|_F(\mathbf{p}^*,\boldsymbol{\epsilon}^*) +  \mu\nabla  h|_F(\mathbf{p}^*,\boldsymbol{\epsilon}^*)$. We thus have that 
\begin{align}
    \epsilon_j^* &= \lambda_1\cos\epsilon_j^*+\lambda_2\sin\epsilon_j^* + \mu, \label{Dtretton}\\
    1 &= \lambda_2 \cos\epsilon_j^*-\lambda_1\sin\epsilon_j^*, \label{Dfemton} \\
    0 &= \lambda_1 + \mu.\label{Dfjorton}
\end{align}
Equations \eqref{Dtretton} and \eqref{Dfemton} hold for $j=1,2,\dots,l$. If we move $\mu$ to the left side in \eqref{Dtretton}, square both \eqref{Dtretton} and \eqref{Dfemton}, and then add them, we get the equation $(\epsilon_j^*-\mu)^2=\lambda_1^2+\lambda_2^2$. Thus there are only two possibilities for the $\epsilon_j^*$s,
\begin{align}
	&\bar\epsilon_1=\mu+\sqrt{\lambda_1^2+\lambda_2^2},\\
	&\bar\epsilon_2=\mu-\sqrt{\lambda_1^2+\lambda_2^2}.
\end{align}
We assume both are present among the $\epsilon_j^*$s and, therefore, are strictly positive. The multiplier $\mu$, and hence $-\lambda_1$, is the arithmetic mean of $\bar\epsilon_1$ and $\bar\epsilon_2$. It follows that
\begin{equation}
    \lambda_2^2
    =(\bar\epsilon_1-\mu)^2-\lambda_1^2 
    =\bigg(\frac{\bar\epsilon_1-\bar\epsilon_2}{2}\bigg)^2-\bigg(\frac{\bar\epsilon_1+\bar\epsilon_2}{2}\bigg)^2 
    =-\bar\epsilon_1\bar\epsilon_2.
\end{equation}
This equation cannot be satisfied for positive $\bar\epsilon_1$ and $\bar\epsilon_2$. Thus, we have reached a contradiction. The conclusion is that only one of $\bar\epsilon_1$ and $\bar\epsilon_2$ is present among the $\epsilon_j^*$s and, therefore, all the $\epsilon_j^*$s have the same value.

\subsection{Conclusion}
Appendices \ref{not dist} and \ref{full dist} show that if $\tau\langle H-\epsilon_0\rangle$ assumes its smallest possible value under the requirement that the state evolves between two states with fidelity $\delta$, then the state is a partly grounded effective qubit. Consequently, the case reduces to that treated in Sec.\ \ref{sec: 2D}, regardless of the dimension of the system. We conclude that $\alpha(n,\delta)=\alpha(\delta)$ and thus that the extended Margolus-Levitin QSL \eqref{eq: the Margolus-Levitin quantum speed limit} is valid in all dimensions.

\section{Geodesics and ruled Seifert surfaces}\label{appendix B}
A curve $\rho_t$ in $\PP$ is a geodesic if its acceleration vanishes, $\nabla_{\dot\rho_t}\dot\rho_t=0$. For the Fubini-Study metric, $\nabla_{\dot\rho_t}\dot\rho_t=[[\ddot\rho_t,\rho_t],\rho_t]$; see, e.g., Ref.\ \cite{HoAlSo2022}. Let $\sigma$ and $\rho$ be two states at a distance $r<\pi/2$ apart. There is a unique shortest unit speed geodesic from $\sigma$ to $\rho$: Choose a unit vector $\ket{\phi}$ over $\sigma$ and let $\ket{\psi}$ be the unit vector over $\rho$ that is in phase with $\ket{\phi}$. Then $\braket{\phi}{\psi}=\cos r$. Define $\ket{w}$ by the condition $\ket{\psi}= \cos r \ket{\phi} + \sin r \ket{w}$, and let $\ket{\phi_t}=\cos t\ket{\phi}+\sin t\ket{w}$. The curve
\begin{equation}\label{seeone}
	\gamma_t
    = \ketbra{\phi_t}{\phi_t} 
	= \cos^2 t\ketbra{\phi}{\phi}+\cos t\sin t\big(\ketbra{\phi}{w}+\ketbra{w}{\phi}\big)
		+\sin^2 t\ketbra{w}{w}, 
  \quad 0\leq t\leq r,
\end{equation}
is the shortest unit speed geodesic that extends from $\sigma$ to $\rho$; straightforward calculations show that $[[\ddot\gamma_t,\gamma_t],\gamma_t]=0$, that $g(\dot\gamma_t,\dot\gamma_t)=1$, and that $\gamma_t$ has the length $r=\dist(\sigma,\rho)$.

\subsection{Optimal evolution curves are sub-Riemannian geodesics}
If the extended Margolus-Levitin QSL is saturated, the state is a partly grounded effective qubit:
\begin{equation}
	\rho = \cos^2 r\ketbra{0}{0} + \cos r\sin r\big(\ketbra{0}{1} + \ketbra{1}{0}\big ) + \sin^2 r\ketbra{1}{1}.
\end{equation}
Here $\ket{0}$ and $\ket{1}$ are eigenvectors of the Hamiltonian with eigenvalues $\epsilon_0$ and $\epsilon_1$, respectively. This state evolves on the geodesic sphere $S(\sigma,r)$ around $\sigma=\ketbra{0}{0}$ with radius $r=\dist(\sigma,\rho)$. The evolution curve is
\begin{equation}
	\rho_t = \cos^2 r\ketbra{0}{0} + \cos r\sin r\big(e^{it(\epsilon_1-\epsilon_0)}\ketbra{0}{1} + e^{-it(\epsilon_1-\epsilon_0)}\ketbra{1}{0}\big) + \sin^2 r\ketbra{1}{1}.
\end{equation}
This curve has the acceleration
\begin{equation}\label{Bee4}
 \nabla_{\dot\rho_t}\dot\rho_t
	= \frac{(\epsilon_1-\epsilon_0)^2}{4} \sin(4r)\Big( \sin(2r) \big( \ketbra{0}{0} - \ketbra{1}{1} \big)
  - \cos(2r) \big( e^{it(\epsilon_1-\epsilon_0)} \ketbra{0}{1} + e^{-it(\epsilon_1-\epsilon_0)} \ketbra{1}{0} \big)\Big).
\end{equation}
The acceleration vanishes identically if and only if $r = \pi/4$. In this case, the state moves along the equator in the projectivization of the span of $\ket{0}$ and $\ket{1}$. Otherwise, the evolution curve is not a geodesic in $\PP$. However, since the acceleration is everywhere perpendicular to $S(\sigma,r)$, the evolution curve is a sub-Riemannian geodesic in $S(\sigma,r)$, that is, a geodesic when considered a curve in $S(\sigma,r)$ with the sub-Riemannian geometry. To see this, fix a $t$ and let $\gamma_s$, $0\leq s\leq r$, be the shortest, arclength-parametrized geodesic from $\sigma$ to $\rho_t$. According to Gauss's lemma \cite{Sa1996}, its velocity vector at $s = r$ is perpendicular to $S(\sigma,r)$. Explicitly, we have that
\begin{equation}\label{Bee5}
	\dot\gamma_r
	=\sin(2r)\big(\ketbra{1}{1}-\ketbra{0}{0}\big)+\cos(2r)\big(e^{it(\epsilon_1-\epsilon_0)}\ketbra{0}{1}+e^{-it(\epsilon_1-\epsilon_0)}\ketbra{1}{0}\big).
\end{equation}
By equations \eqref{Bee4} and \eqref{Bee5}, $\nabla_{\dot\rho_t}\dot\rho_t=-\frac{1}{4}(\epsilon_1-\epsilon_0)^2\sin(4r)\dot\gamma_r$, which implies that $\nabla_{\dot\rho_t}\dot\rho_t$ is perpendicular to $S(\sigma,r)$ at $\rho_t$. Consequently, $\rho_t$ is a sub-Riemannian geodesic.

\subsection{Ruled Seifert surfaces}
Let $\rho_t$, $0\leq t\leq\tau$, be a curve in $\PP$ which for each $t$ is at the distance $r<\pi/2$ from $\sigma$. We can construct a Seifert surface for its $\sigma$ closure in the following way. Let $\ket{\phi}$ be a lift of $\sigma$, let $\ket{\psi_t}$ be the lift of $\rho_t$ that is in phase with $\ket{\phi}$, define $\ket{w_t}$ by the condition $\ket{\psi_t}= \cos r \ket{\phi} + \sin r \ket{w_t}$, and set $\ket{\psi_{s,t}}=\cos s\ket{\phi}+\sin s\ket{w_t}$ for $0\leq s\leq r$ and $0\leq t\leq \tau$. Then $\Sigma(s,t)=\ketbra{\psi_{s,t}}{\psi_{s,t}}$ is a Seifert surface for $\rho_t^\sigma$ consisting of geodesics starting from $\sigma$ and ending at points on $\rho_t$. 

\section{Extreme values of $\JJ^\sigma_{r,\delta}$}\label{appendix C}
Let $\sigma$ be an arbitrary state and $\Gamma(\sigma,r,\delta)$ be the set of smooth curves in $S(\sigma,r)$ that stretches between two states with fidelity $\delta$. Consider the functional $\JJ^\sigma_{r,\delta}$ on $\Gamma(\sigma,r,\delta)$ defined as
\begin{equation}
    \JJ^\sigma_{r,\delta}[\rho_t]
    = \int_{\rho_t}A^\sigma.
\end{equation}
We calculate the extreme values of $\JJ^\sigma_{r,\delta}$. 

Fix an arbitrary lift $\ket{\phi}$ of $\sigma$ and recall that $Y_{\ket{\phi}}$ is the hypersurface in $\SS$ consisting of all vectors in phase with $\ket{\phi}$. Let $\ket{\phi}$ be the first vector in an orthonormal basis $\ket{\phi},\ket{1},\dots,\ket{n-1}$ for $\HH$. Every lift $\ket{\psi_t}$ of a curve in $S(\sigma,r)$ to $Y_{\ket{\phi}}$ has a decomposition of the form $\ket{\psi_t}=\cos r\ket{\phi} + \sin r \ket{w_t}$, where
\begin{equation}
	\ket{w_t}
	=\sum_{j=1}^{n-1} (u_{j;t}+iv_{j;t})\ket{j}
\end{equation}
for some real-valued functions $u_{j;t}$ and $v_{j;t}$ satisfying
\begin{equation}
	\sum_{j=1}^{n-1} (u_{j;t}^2+v_{j;t}^2)=1.
\end{equation}
Conversely, any such curve in $Y_{\ket{\phi}}$ projects to a curve in $S(\sigma,r)$. Let $\rho_t=\ketbra{\psi_t}{\psi_t}$ and assume that $0\leq t\leq \tau$. Then
\begin{equation}
	\JJ^\sigma_{r,\delta}[\rho_t]
	=\sum_{j=1}^{n-1} \int_0^\tau\dt \,(u_{j;t}\dot v_{j;t} - \dot u_{j;t} v_{j;t}).
\end{equation}

Suppose that the fidelity between $\rho_0$ and $\rho_\tau$ is $\delta$ and that $\rho_t$ is extremal for $\JJ^\sigma_{r,\delta}$. The $u_{j;t}$s and $v_{j;t}$s then satisfy the Euler-Lagrange equations for the augmented Lagrangian
\begin{equation}
    L(u_j,\dot u_j, v_j,\dot v_j,\lambda)
    =\sum_{j=1}^{n-1} (u_j\dot v_j-\dot u_j v_j)
    -\lambda\bigg( \sum_{j=1}^{n-1}(u^2_j+v^2_j)
    -1\bigg).
\end{equation}
The Euler-Lagrange equations read
\begin{align}
	&\dot u_j=-\lambda v_j,\\	
	&\dot v_j = \lambda u_j,\\
	&\sum_{j=1}^{n-1} (u^2_j+v^2_j)=1.
\end{align}
These equations imply that $u_{j;t} + iv_{j;t} = e^{i\Lambda_t}(u_{j;0} + iv_{j;0})$, with $\Lambda_t$ being the integral of the Lagrange multiplier,
\begin{equation} 
    \Lambda_t=\int_0^t\d s\,\lambda_s.
\end{equation}
Thus, $\ket{\psi_t}=\cos r\,\ket{\phi} + \sin r\, e^{i\Lambda_t}\ket{w_0}$. That the fidelity between $\rho_0$ and $\rho_\tau$ is $\delta$ translates to
\begin{equation}
    \delta = 1+\frac{1}{2}\sin^22r(\cos\Lambda_\tau-1).
\end{equation}
We conclude that
\begin{equation}
    \Lambda_\tau
    =\pm\arccos\left(1-\frac{2(1-\delta)}{\sin^22r}\right)\mod 2\pi
\end{equation}
and, hence, that 
\begin{equation}
	\JJ^\sigma_{r,\delta}[\rho_t]
	=\pm\sin^2r\,\arccos\bigg(1-\frac{2(1-\delta)}{\sin^22r}\bigg) \mod 2\pi.
\end{equation}

\twocolumngrid


\begin{thebibliography}{99}

\bibitem{Fr2016}
M. R. Frey,
Quantum speed limits---primer, perspectives, and potential future directions,
\href{https://doi.org/10.1007/s11128-016-1405-x}{Quantum Inf. Process. {15}, 3919 (2016)}.

\bibitem{DeCa2017} 
S. Deffner and S. Campbell,
Quantum speed limits: From Heisenberg's uncertainty principle to optimal quantum control, 
\href{https://doi.org/10.1088/1751-8121/aa86c6}{J. Phys. A: Math. Theor. {50}, 453001 (2017)}.

\bibitem{De2020} 
S. Deffner,
Quantum speed limits and the maximal rate of information production,
\href{https://doi.org/10.1103/PhysRevResearch.2.013161}{Phys. Rev. Res. {2}, 013161 (2020)}

\bibitem{PiCiCeAdS-P2016} 
 D. P. Pires, M. Cianciaruso, L. C. C\'eleri, G. Adesso, and D. O. Soares-Pinto,
Generalized geometric quantum speed limits,
\href{https://doi.org/10.1103/PhysRevX.6.021031}{Phys. Rev. X {6}, 021031 (2016)}

\bibitem{dCEgPlHu2013}
A. del Campo, I. L. Egusquiza, M. B. Plenio, and S. F. Huelga,
Quantum speed limits in open system dynamics, 
\href{https://doi.org/10.1103/PhysRevLett.110.050403}{Phys. Rev. Lett. {110}, 050403 (2013)}.

\bibitem{MaTa1945}
L. Mandelstam and I. Tamm, 
The uncertainty relation between energy and time in non-relativistic quantum mechanics, 
\href{https://doi.org/10.1007/978-3-642-74626-0_8}{J. Phys. (USSR) {9}, 249 (1945)}.

\bibitem{AnAh1990}
J. Anandan and Y. Aharonov,
Geometry of quantum evolution,
\href{https://doi.org/10.1103/PhysRevLett.65.1697}{Phys. Rev. Lett. {65}, 1697 (1990)}.

\bibitem{HoAlSo2022}
N. H\"ornedal, D. Allan, and O. S\"onnerborn,
Extensions of the Mandelstam–Tamm quantum speed limit to systems in mixed states,
\href{https://doi.org/10.1088/1367-2630/ac688a}{New J. Phys. {24}, 055004 (2022)}.

\bibitem{MaLe1998}
N. Margolus and L. B. Levitin,
The maximum speed of dynamical evolution,
\href{https://doi.org/10.1016/S0167-2789(98)00054-2}{Phys. D {120}, 188 (1998)}. 

\bibitem{GiLlMa2003}
V. Giovannetti, S. Lloyd, and L. Maccone,
Quantum limits to dynamical evolution,
\href{https://doi.org/10.1103/PhysRevA.67.052109}{Phys. Rev. A {67}, 052109 (2003)}.

\bibitem{TaEsDaMa-Fi2013}
M. M. Taddei, B. M. Escher, L. Davidovich, and R. L. de Matos Filho,
Quantum speed limit for physical processes, 
\href{https://doi.org/10.1103/PhysRevLett.110.050402}{Phys. Rev. Lett. {110}, 050402 (2013)}.

\bibitem{Ta2014}
M. M. Taddei,
Quantum Speed Limits for General Physical Processes, 
Ph.D. thesis, Universidade Federal do Rio de Janeiro, 2014 (unpublished), \href{https://doi.org/10.48550/arXiv.1407.4343}{arXiv:1407.4343}.

\bibitem{AhAn1987}
Y. Aharonov and J. Anandan, 
Phase change during a cyclic quantum evolution, 
\href{https://doi.org/10.1103/PhysRevLett.58.1593}{Phys. Rev. Lett. 58, 1593 (1987)}

\bibitem{HoSo2023}
N. H\"ornedal and O. S\"onnerborn,
Closed systems refuting quantum-speed-limit hypotheses,
\href{https://doi.org/10.1103/PhysRevA.108.052421}{Phys. Rev. A {108}, 052421 (2023)}.

\bibitem{SoBjTsTr1999}
J. S\"oderholm, G. Bj\"ork, T. Tsegaye, and A. Trifonov, 
States that minimize the evolution time to become an orthogonal state,
\href{https://doi.org/10.1103/PhysRevA.59.1788}{Phys. Rev. A {59}, 1788 (1999)}.

\bibitem{LeTo2009}
L. B. Levitin and T. Toffoli,
Fundamental limit on the rate of quantum dynamics: The unified bound is tight,
\href{https://doi.org/10.1103/PhysRevLett.103.160502}{Phys. Rev. Lett. {103}, 160502 (2009)}.

\bibitem{MuSi1993}
N. Mukunda and R. Simon,
Quantum kinematic approach to the geometric phase. I. General formalism,
\href{https://doi.org/10.1006/aphy.1993.1093}{Ann. Phys. {228}, 205 (1993)}.

\bibitem{Br1993}
G. E. Bredon,
\textit{Topology and Geometry, Graduate Texts in Mathematics} (Springer, New York, 1993).

\bibitem{HoSo2023c}
N. H\"ornedal and O. S\"onnerborn,
Tight lower bounds on the time it takes to generate a geometric phase, 
\href{https://doi.org/10.1088/1402-4896/acf8a2}{Phys. Scr. {98}, 105108 (2023)}.

\bibitem{NeAlSa2022}
G. Ness, A. Alberti, and Y. Sagi,
Quantum speed limit for states with a bounded energy spectrum,
\href{https://doi.org/10.1103/PhysRevLett.129.140403}{Phys. Rev. Lett. {129}, 140403 (2022)}.

\bibitem{Br2003}
D. C. Brody,
Elementary derivation for passage times,
\href{https://doi.org/10.1088/0305-4470/36/20/314}{J. Phys. A: Math. Gen. {36}, 5587 (2003)}.

\bibitem{AnHe2014}
O. Andersson and H. Heydari,
Quantum speed limits and optimal Hamiltonians for driven systems in mixed states,
\href{https://doi.org/10.1088/1751-8113/47/21/215301}{J. Phys. A: Math. Theor. {47}, 215301 (2014)}.

\bibitem{An2018}
O. Andersson,
Holonomy in Quantum Information Geometry, Ph. Lic. thesis, Stockholm University, 2018 (unpublished), \href{https://arxiv.org/abs/1910.08140}{arXiv:1910.08140}.

\bibitem{AnAl2004}
M. Andrecut and M. K. Ali,
The adiabatic analogue of the Margolus–Levitin theorem,
\href{https://doi.org/10.1088/0305-4470/37/15/L01}{J. Phys. A: Math. Gen. {37}, L157 (2004)}.

\bibitem{Uh1976}
A. Uhlmann,
The ``transition probability'' in the state space of a $\ast$-algebra,
\href{https://doi.org/10.1016/0034-4877(76)90060-4}{Rep. Math. Phys. {9}, 273 (1976)}.

\bibitem{Uh1995}
A. Uhlmann,
Geometric phases and related structures,
\href{https://doi.org/10.1016/0034-4877(96)83640-8}{Rep. Math. Phys. {36}, 461 (1995)}.

\bibitem{SjPaEkAnErOiVe2000}
E. Sj\"oqvist, A. K. Pati, A. Ekert, J. S. Anandan, M. Ericsson, D. K. L. Oi, and V. Vedral,
Geometric phases for mixed states in interferometry,
\href{https://doi.org/10.1103/PhysRevLett.85.2845}{Phys. Rev. Lett. {85}, 2845 (2000)}

\bibitem{AnHe2015}
O. Andersson and H. Heydari,
A symmetry approach to geometric phase for quantum ensembles,
\href{https://doi.org/10.1088/1751-8113/48/48/485302}{J. Phys. A: Math. Theor. {48}, 485302 (2015)}.

\bibitem{AnHe2013}
A. Uhlmann,
Operational geometric phase for mixed quantum states,
\href{https://doi.org/10.1088/1367-2630/15/5/053006}{New J. Phys. {15}, 053006 (2013)}.

\bibitem{Sa1996}
T. Sakai,
\textit{Riemannian Geometry, Translations of Mathematical Monographs} (American Mathematical Society, Providence, Rhode Island, 1996), Vol.\ 149. 

\end{thebibliography}
\end{document}